\documentclass[aps,prd,nofootinbib,twocolumn,reprint,superscriptaddress,longbibliography,showpacs,showkeys]{revtex4}
\pdfoutput=1
\usepackage{amsmath}
\usepackage{amssymb}
\usepackage{amsfonts}
\usepackage{amsthm}
\usepackage{mathrsfs}
\usepackage{natbib}
\usepackage{latexsym}
\usepackage{graphicx}
\usepackage{lipsum}
\usepackage{dsfont}
\usepackage{txfonts}
\usepackage{rotating}
\usepackage[normalem]{ulem}
\usepackage{wasysym}
\usepackage{multirow}
\usepackage{hhline}
\usepackage{graphicx}
\usepackage{hyperref}
\hypersetup{colorlinks = true}
\usepackage[dvipsnames,usenames]{color}
\usepackage{bm}
\usepackage{appendix}
\usepackage{acronym}
\usepackage{comment}
\usepackage{enumitem}
\usepackage{bigdelim}
\usepackage{algorithm}
\usepackage[noend]{algpseudocode}
\usepackage{qcircuit}
\usepackage{graphicx}
\usepackage{subfig}
\usepackage{array}
\usepackage{academicons}
\usepackage{xcolor}
\usepackage{cancel}
\usepackage{tikz}
\usepackage{framed}

\newcommand{\bra}[1]{\langle #1|}
\newcommand{\ket}[1]{|#1\rangle}
\newcommand{\braket}[2]{\langle #1|#2\rangle}

\definecolor{lime}{HTML}{A6CE39}
\DeclareRobustCommand{\orcidicon}{%
	\begin{tikzpicture}
	\draw[lime, fill=lime] (0,0) 
	circle [radius=0.16] 
	node[white] {{\fontfamily{qag}\selectfont \tiny ID}};
	\draw[white, fill=white] (-0.0625,0.095) 
	circle [radius=0.007];
	\end{tikzpicture}
	\hspace{-2mm}
}

\foreach \x in {A, ..., Z}{%
	\expandafter\xdef\csname orcid\x\endcsname{\noexpand\href{https://orcid.org/\csname orcidauthor\x\endcsname}{\noexpand\orcidicon}}
}

\begin{document}

\pacs{03.67.Ac; 04.30.-w; 07.05.Kf}
\keywords{Quantum algorithm, matched filtering, Grover's algorithm, gravitational waves, continuous waves, data analysis}

\title{A quantum algorithm for gravitational wave matched filtering}

% NOTE TO EVERYBODY: we should decide on whether to make
% names full or initialized -- GDM
\author{Sijia Gao\orcidA{}}
\email{s.gao.2@research.gla.ac.uk}
\affiliation{SUPA, School of Physics and Astronomy, University of Glasgow, Glasgow G12 8QQ, United Kingdom}
\author{Fergus Hayes\orcidB{}}
\email{f.hayes.1@research.gla.ac.uk}
\affiliation{SUPA, School of Physics and Astronomy, University of Glasgow, Glasgow G12 8QQ, United Kingdom}
\author{Sarah Croke\orcidC{}}
\affiliation{SUPA, School of Physics and Astronomy, University of Glasgow, Glasgow G12 8QQ, United Kingdom}
\author{Chris Messenger\orcidD{}}
\affiliation{SUPA, School of Physics and Astronomy, University of Glasgow, Glasgow G12 8QQ, United Kingdom}
\author{John Veitch\orcidE{}}
\affiliation{SUPA, School of Physics and Astronomy, University of Glasgow, Glasgow G12 8QQ, United Kingdom}

\date{\today}
%\date{\commitDATE\\\mbox{\small{\commitID} \commitSTATUS}\\}

\begin{abstract}
Quantum computational devices, currently under development, have the potential to accelerate data analysis techniques beyond the ability of any classical algorithm.
We propose the application of a quantum algorithm for the detection of unknown signals in noisy data.
We apply Grover's algorithm to matched-filtering, a signal processing technique that compares data to a number of candidate signal templates.
%to identify template signals that fit the data with a high signal-to-noise ratio.
In comparison to the classical method, this provides a speed-up proportional to the square-root of the number of templates, which would make possible otherwise intractable searches.
%This provides a square-root speed-up in comparison to the classical method, and promises to make possible otherwise intractable searches. 
We demonstrate both a proof-of-principle quantum circuit implementation, and a simulation of the algorithm's application to the detection of the first gravitational wave signal GW150914. 
We discuss the time complexity and space requirements of our algorithm as well as its implications for the currently computationally-limited searches for continuous gravitational waves.

% However, the detection of certain astrophysical sources is made extremely difficult by the prohibitive computational power required to match potential signals hidden in data against huge numbers of possible template waveforms. Quantum computational devices, currently under development, have the potential to accelerate this type of analysis beyond the ability of any classical algorithm. We investigate the application of a quantum algorithm for fast signal detection. We apply Grover's algorithm to matched-filtering searches by identifying templates that return a significant signal-to-noise ratio. This provides a square-root speed-up in comparison to the classical method, and promises to make possible otherwise intractable searches. We demonstrate the application of the algorithm both with a quantum simulator and when applied to the detection of the first gravitational wave signal GW150914. We discuss the time complexity and space requirements of our algorithm as well as its implications for the future discovery of continuous waves.\fergus{I have made changes to the abstract given old comments and added a final comment on continuous waves. We do still need to revise this soon.}
\end{abstract} 

\maketitle

\acrodef{NS}[NS]{Neutron Star}
\acrodef{GW}[GW]{gravitational-wave}
\acrodef{SNR}[SNR]{signal-to-noise ratio}
\acrodef{qRAM}[qRAM]{quantum random access memory}
\acrodef{PSD}[PSD]{power spectral density}
\acrodef{FFT}[FFT]{fast Fourier transform}
\acrodef{CBC}[CBC]{compact binary coalescence}
\acrodef{NP}[NP]{non-deterministic polynomial-time}
\acrodef{QFT}[QFT]{quantum Fourier tranform}
\acrodef{SSB}[SSB]{solar system barycentre}

%%%%%%%%%%%%%%%%%%%%%%%%%%%%%%%%%%%%%%%%%%%%%%%%
%%%%%%%%%%%%%%%%%%%%%%%%%%%%%%%%%%%%%%%%%%%%%%%%
\section{Introduction}\label{sec:intro}

%TODO: Write snappy opening paragraph

Quantum computing holds enormous potential for computational speed-up of certain tasks, offering the possibility of solving classically intractable problems, in particular in quantum chemistry and many body physics \cite{review2020qchemRMP,review2020ChemRev}. The technology has seen rapid development in the last few years, resulting in processors with 50-100 qubits, and the first demonstrations of clear quantum advantage over classical computation~\cite{arute2019quantum,zhong2020quantum}. Quantum algorithms (see \cite{montanaro2016review} for an accessible overview) are being explored for more and more fields of endeavour: for example finance \cite{orus2019quantum}, quantum simulation \cite{georgescu2014quantum}, particle physics \cite{blance2021quantum,magano2021investigating}, machine learning \cite{Dunjko2018review,Biamonte2017review}, and as the technology matures and a new generation of software developers adopt quantum programming languages, it may be anticipated that new and unexpected applications will be discovered. A particularly versatile quantum sub-routine is Grover's search algorithm \cite{grover1996fast}, which finds a marked solution in a large unstructured database. Grover's algorithm, one of the earliest proposed quantum algorithms, provides a square-root speed up over classical search. This is less dramatic than the exponential speed-up promised by e.g. Shor's algorithm \cite{shor1995factoring}, but can nevertheless provide a significant practical advantage for problems with a large search space. By defining the search space and conditions for a desired solution, Grover's algorithm may be applied to any computational problem with limited structure, and has found use in minimum finding \cite{durr1996quantum}; clustering and nearest neighbour algorithms for supervised and unsupervised learning \cite{aimeur2013quantum,wiebe2015quantum}; and pattern matching \cite{ramesh2003string,montanaro2017quantum,niroula2021quantum}, to name but a few. In this paper we propose the use of Grover's search in quantum algorithms for matched filtering, with applications in gravitational wave astronomy. These algorithms inherit the square root speed up of Grover's search algorithm, an improvement which could enable gravitational wave searches currently intractable with state-of-the-art classical techniques.

Matched filtering is a signal processing technique in which an exhaustive search is performed over a bank of templates to find the template that when correlated with the data returns the highest detection statistic, making it a natural candidate for a quantum speed-up through Grover's algorithm. In gravitational wave matched-filtering a geometric definition of distance within the parameter space is defined based on the relative loss in \ac{SNR} between a template and a potential signal. The required distribution of the templates in the search space are chosen so that the distance (or overlap) between adjacent templates is constant throughout the space. Depending on the specific data analysis problem, the number of templates can range up to ${\sim}10^{12}$~\cite{2019ApJ...875..122A} resulting in a total computational time of ${\sim}10^6$ CPU hours. The spacing of templates in the parameter space determines the efficiency of the search, but also the overall number of templates, and the sensitivity of searches for certain classes of signals (e.g. continuous wave sources) is currently computationally limited. Thus even a modest square-root speed-up could enable the detection of signals which would be infeasible with classical techniques.

%In gravitational wave matched-filtering a geometric definition of distance within the parameter space is defined based on the relative loss in \ac{SNR} between a template and a potential signal. The required distribution of the templates in the search space are chosen so that the distance (or overlap) between adjacent templates is constant throughout the space. Once a bank of templates is constructed, an exhaustive search is performed over the bank to find the template, that when correlated with the data, returns the highest detection statistic. Depending on the specific data analysis problem, the number of templates can range up to ${\sim}10^{12}$~\cite{2019ApJ...875..122A} resulting in a total computational time of ${\sim}10^6$ CPU hours. Gravitational wave astronomy thus provides a natural, but so far unexplored application for which scalable, fault tolerant quantum computers could offer a tangible benefit. 

Key to our proposed algorithms is the fact that the potential signals in gravitational wave astronomy are well-modelled by general relativity, and the templates may be readily computed as part of the matching procedure. This eliminates the need to pre-load the database into \ac{qRAM} \cite{giovannetti2008qRAM}, and thus avoids hidden complexity associated with this loading step, as well as doubts about the experimental feasibility of constructing \ac{qRAM}~\cite{aaronson2015,Preskill2018,ciliberto2018,tang2021quantum}. The presented algorithms may be applied to any matched filtering problem in which the required templates may be efficiently computed, although we focus here on the application to gravitational wave detection. A range of quantum algorithms for data processing and more general learning tasks exist in the literature (e.g. \cite{Dunjko2018review,Biamonte2017review,aimeur2013quantum,wiebe2015quantum,schutzhold2003pattern,lloyd2013quantum,Lloyd2014qpca,schuld2014quantum,wiebe2012quantum,amin2018quantum,brandao2019quantum}), but to our knowledge this is the first proposal for an application to matched filtering, a widely used signal processing technique \cite{helstrom1968statistical,2008arXiv0804.1161S}. Most closely related to our work are existing algorithms for pattern matching \cite{ramesh2003string,montanaro2017quantum,niroula2021quantum}, which search for an exact or approximate match for a specified pattern (bit string) within a larger dataset; these however require the data and pattern to be loaded into memory, which would have prohibitive space requirements in the case considered here. Alternatively, algorithms for quantum template matching were first proposed almost twenty years ago \cite{sasaki2001,sasaki2002}, in which optimal strategies for determining the closest matching template are given. These rely on generalised quantum measurements with one outcome for each possible template; translating a gravitational wave template bank into such a measurement is not trivial for the simplest cases, and likely infeasible for the more interesting cases. A related task in the literature is estimating the overlap between quantum states, provided a number of copies of each~\cite{buhrman2001,fanizza2020}. In gravitational wave data analysis however the number of templates is by far the largest parameter, and such an approach does not obviously offer an advantage.

Although current state-of-the-art quantum processors are still too small and error-prone for many applications of interest, there is much effort concentrated around developing applications for so-called noisy intermediate-scale quantum (NISQ) devices~\cite{Preskill2018}, with quantum machine learning being one promising area~\cite{ciliberto2018,Biamonte2017review,Dunjko2018review}. The next technological hurdle will be to implement error-correction, and this comes with an overhead in the number of physical qubits required in order to produce a smaller number of error-corrected logical qubits \cite{preskill1997,gottesman2009}. In the longer run fully scalable, fault tolerant devices will be required for universal quantum computation, and to run algorithms such as Shor's famous factoring algorithm \cite{shor1995factoring}. At this point further applications in machine learning, pattern matching, and data processing may be expected, to which we now add matched filtering for gravitational wave data analysis.

In the remainder of the paper we show how to employ Grover's algorithm and its extension to quantum counting to perform quantum matched filtering. We choose a digital encoding for the data and templates, that is, each is encoded as classical bits in the computational basis, and explicitly construct a quantum oracle which returns whether a template matches with the data above a given threshold. We present two algorithms demonstrating the application of quantum counting to matched filtering; the first determines whether there is at least one matching template and provides an estimate to their number; the second returns matching templates. We require only that there is an efficient classical algorithm to generate the templates from an index into the considered set of parameters, and to perform template matching. %Our algorithm does not require \ac{qRAM}~\cite{giovannetti2008qRAM}, and thus avoids hidden complexity associated with loading data and templates into \ac{qRAM}. The algorithms inherits the square-root speed-up of Grover's algorithm over a classical brute force exhaustive search over templates. 
We discuss the complexity of our algorithms compared to classical techniques, and the implications for gravitational wave data analysis. We go beyond an asymptotic analysis to compare the approximate number of matching calculations needed in the classical and quantum algorithms for particular match-filtering problems and defined performance requirements, showing orders of magnitude of difference between the quantum and classical algorithms.

Throughout it is our aim to present our ideas in a form accessible to both the gravitational wave and quantum computing communities. Thus we provide some background and details to each which will be well-known to experts within each field, but may be unfamiliar to the other subset of the intended audience. In Sec.~\ref{sec:background} we review gravitational waves, matched-filtering, Grover's algorithm and quantum counting. Following this we present our algorithm in Sec.~\ref{sec:psuedocode}. We give an implementation on IBM's Qiskit platform~\cite{Qiskit} in Sec.~\ref{sec:qizkitexample}, and an analysis of the application to the detection of the first gravitational wave detected, GW150914, in Sec.~\ref{sec:cbcexample}. We detail the potential speed-up provided by our algorithm for matched filtering applied to continuous waves in Sec~\ref{sec:cw_example} and discuss the implications to their discovery. We conclude with a discussion of the implications of our work, and suggest directions for further study. We also include an introduction to quantum computing concepts in Appendix.~\ref{sec:quantumgates} and some of the mathematical details in Appendix.~\ref{sec:Pt0}. 

%%%%%%%%%%%%%%%%%%%%%%%%%%%%%%%%%%%%%%%%%%%%%%%%
%%%%%%%%%%%%%%%%%%%%%%%%%%%%%%%%%%%%%%%%%%%%%%%%
%%%%%%%%%%%%%%%%%%%%%%%%%%%%%%%%%%%%%%%%%%%%%%%%
%%%%%%%%%%%%%%%%%%%%%%%%%%%%%%%%%%%%%%%%%%%%%%%%
\section{Background}\label{sec:background}

\subsection{Gravitational wave searches}
% GW introduction
The detection of gravitational waves from the merger of compact binary systems is now a regular occurrence. Since the first detection of the binary black hole merger, known as GW150914~\cite{2016PhRvL.116f1102A}, the Advanced LIGO and Advanced Virgo detectors have detected signals from 50 such systems including two binary neutron star systems~\cite{2020arXiv201014527A}.  The individual detections, and the population as a whole, allow us to infer properties of gravitational wave sources including the nature of extreme matter constituting neutron stars~\cite{2018PhRvL.121p1101A}, set stringent constraints on the accuracy of general relativity~\cite{2020arXiv201014529T}, resolve the mystery of the origin of short gamma-ray bursts~\cite{2017ApJ...848L..13A}, probe the formation history of compact objects~\cite{2020arXiv201014533T}, and make new measurements on cosmological parameters independent of the cosmic distance ladder~\cite{2019arXiv190806060T}.

% other sources yet to detect
While searches are ongoing for continuously emitted gravitational waves, supernovae and unmodelled burst sources, and the astrophysical and cosmological stochastic backgrounds, as yet only signals from compact binary coalescences have been detected. However, as the advanced gravitational wave detectors~\cite{Aasi:2013wya, TheLIGOScientific:2014jea, TheVirgo:2014hva} increase in sensitivity and additional detectors join the global network~\cite{2020arXiv200505574A,LIGOIndia} our reach into the universe grows. With sensitivity to greater cosmic distances the rate of detections will grow and other intrinsically weaker classes of signal (e.g., continuous gravitational waves) will become detectable (see~\cite{2020arXiv200714251T} for the most recent results from searches for the known millisecond pulsars).       
% computationally difficult
% Describe scales of problem: CBC -> Coherent CBC -> CW

The compact binary and continuous gravitational wave sources are subject to a matched-filtering search approach~\cite{1996PhRvD..53.3033B,1996PhRvD..53.6749O,1999PhRvD..60b2002O,1998PhRvD..57.2101B,Allen:2005fk}. This is motivated by the fact that these sources are very well modelled by general relativity. For the transient compact binary signals, template waveforms are obtained through post-Newtonian expansion of the orbital dynamics and calibrated against numerical relativity simulations for the merger and ring-down phase~\cite{2016PhRvD..93d4006H,2017PhRvD..95d4028B}. The continuous wave case is somewhat simpler since the waveform is expected to be a weak sinusoid generated by rotating neutron stars with non-zero mass quadrupole moments. Such sources will exhibit slowly varying Doppler modulation of the frequency due to the motion of the detector relative to the source, combined with amplitude modulation produced by the antenna response of the detector as the Earth rotates~\cite{1998PhRvD..58f3001J}.

% more computational costs
An additional continuous wave problem is that of searching for signals from sources that reside in binary systems. This leads to an additional dramatic increase in parameter space volume and the corresponding numbers of templates~\cite{2001PhRvD..63l2001D,2011PhRvD..84h3003M,2015PhRvD..91j2003L}. When comparing the compact binary and continuous wave cases, the relative size of the search spaces, and hence the number of required templates, is typically much greater for the continuous wave case~\cite{2019ApJ...875..122A}. In fact, the number of templates required for a fully coherent analysis for a continuous wave source of unknown sky location, frequency, and first frequency time derivative (representing the slow drift in the intrinsic spin of the source), makes such a search completely infeasible. Searches such as these are computationally limited in their sensitivity, and so less sensitive but tractable semi-coherent approaches are applied. Such schemes subdivide the data in either time or frequency space, analyse each part separately and then combine the results in such a way as to ignore the signal phase coherence between segments, significantly reducing the computational cost at the expense of sensitivity. To a lesser extent there are computational limitations for the compact binary searches when extending the search space to precessing systems~\cite{2016PhRvD..94b4012H} and a coherent analysis between different detectors~\cite{2016PhRvD..93f4004M}.

\subsection{Matched filtering}\label{sec:mf}

Matched filtering is a signal processing technique used to maximise the \ac{SNR} by correlating a signal template with measured data. It is the optimal method for detecting a known signal buried in Gaussian noise~\cite{helstrom1968statistical} and is close to optimal for the case of searching over a collection of possible templates~\cite{2008arXiv0804.1161S}. For the derivation of a matched filter, consider the detector output time-series to be $h(t)$, defined:
\begin{equation}\label{equ:strain}
h(t) = s(t) + n(t),
\end{equation}
where $s(t)$ is the signal which is added to some noise $n(t)$. Now consider a linear filter $q(t)$ that is applied to the data in the form of an inner product. Assuming the signal has some finite duration, this can be written in the frequency domain denoted $\tilde{\boldsymbol{\cdot}}$ as:
\begin{equation}\label{equ:dotsnr}
\begin{split}
q\cdot h =& \int_{-\infty}^{\infty} \tilde{q}^{\ast}(f)\tilde{h}(f)\,df\\
=& \int_{-\infty}^{\infty} \tilde{q}^{\ast}(f)\tilde{s}(f)\,df + \int_{-\infty}^{\infty} \tilde{q}^{\ast}(f)\tilde{n}(f)\,df.
\end{split}
\end{equation}
It is evident that $q$ should be chosen as to maximize the inner product with the signal whilst minimizing the expected inner product with the noise. We can define the optimal \ac{SNR} after applying the linear filter terms for the case of zero-mean noise using: 
\begin{equation}\label{equ:insnr}
\begin{split}
\text{SNR}^{2} = & \frac{\left|\int_{-\infty}^{\infty} \tilde{q}^{\ast}(f)\tilde{s}(f)\,df\right|^{2}}
{\text{E}\left[\left|\int_{-\infty}^{\infty} \tilde{q}^{\ast}(f)\tilde{n}(f)\,df\right|^{2}\right]}\\
= & 2\frac{\left|\int_{-\infty}^{\infty} \left(S_{n}^{1/2}(|f|)\tilde{q}(f)\right)^{\ast}\left(S_{n}^{-1/2}(|f|)\tilde{s}(f)\right)df\right|^{2}}
{\int_{-\infty}^{\infty}S_{n}(|f|)|\tilde{q}(f)|^{2}df},
\end{split}
\end{equation}
where $E[\ldots]$ denotes an expection value over noise realisations, and $S_n$ is the single-sided noise \ac{PSD} defined here as: 
\begin{equation}
\frac{1}{2}S_n(|f|)\delta(f-f') = \text{E}\left[\hat{n}(f)\hat{n}^{\ast}(f')\right],
\end{equation}
where $\delta$ is the Dirac delta function. 
This allows for an upper limit to be placed on the \ac{SNR} using the Cauchy-Schwarz inequality, constraining it to
\begin{equation}
\text{SNR}^{2} \le 2\int_{-\infty}^{\infty}S_{n}^{-1}(|f|)|\tilde{s}(f)|^{2}df.
\end{equation}
This upper bound is achieved for Eq.~\ref{equ:insnr} when the template is proportional to the noise-weighted signal $\tilde{s}(f)/S_{n}(f)$. By further applying the constraint that
\begin{equation}\label{equ:snrnorm}
\text{E}\left[\left|\int_{-\infty}^{\infty} \tilde{q}^{\ast}(f)\tilde{n}(f)df\right|^{2}\right] = 1
\end{equation}
gives the constant of proportionality and allows us to define the normalised optimal template:
\begin{equation}\label{equ:optsnr}
\tilde{Q}(f) = \left(\int^{\infty}_{0}S_{n}^{-1}(f)|\tilde{s}(f)|^{2}df\right)^{-1/2}\tilde{s}(f).
\end{equation}

Let us define $\rho(t)$ as the matched filter \ac{SNR} that is determined by applying Eq.~\ref{equ:dotsnr} across $h(t)$ using the optimal template from Eq.~\ref{equ:optsnr}. The inner product in Eq.~\ref{equ:dotsnr} can be applied across signal arrival times by instead considering a convolution, resulting in an additional phase component in the definition of the \ac{SNR}. The matched filter \ac{SNR} can be maximised over the phase at the time of coalescence $\phi_{0}$ by constructing a complex normalised template $\tilde{Q}_{c}(f)$ defined as
\begin{equation}
    \tilde{Q}_{c}(f) = \tilde{Q}_{\phi_{0}=0}(f) + i\tilde{Q}_{\phi_{0}=\pi/4}(f)
\end{equation}
so that the matched filter \ac{SNR} is calculated from the modulus of Eq.~\ref{equ:dotsnr}:
\begin{equation}\label{equ:contpsd}
    \begin{aligned}
    \rho(t) & = & \left|\int^{\infty}_{-\infty}\frac{\tilde{Q}^{\ast}_{c}(f)\tilde{h}(f)}{S_{n}(|f|)}e^{2\pi i t f}df\right|\\
    & = & 2 \left|\int^{\infty}_{0}\frac{\tilde{Q}^{\ast}_{c}(f)\tilde{h}(f)}{S_{n}(f)}e^{2\pi i t f}df\right|.
    \end{aligned}
\end{equation}
For discretised time-series data of $M$ time steps separated by $\Delta t$, $\rho$ as a function of the template and data time offset $t_j$ becomes
\begin{equation}\label{equ:discretesnr}
\rho(t_{j}) = \frac{2}{M\Delta t}\left|\sum^{(M-1)/2}_{k=1}\frac{\tilde{Q}^{\ast}_{c}(f_{k})\tilde{h}(f_{k})}{S_{n}(f_{k})}e^{2\pi ijk/M}\right|.
\end{equation}

The calculation of $\rho$ across all $M$ time steps involves the inverse Fourier transform of the product of the frequency domain signal and template, which has a cost of $O(M^{2})$. 
This process can therefore benefit in computational efficiency via the use of the (classical) \ac{FFT} algorithm, which has a computational cost of $O(M\log M)$~\cite{cooley1967historical}.

For signal detection, the parameter space of interest is discretised and a list of waveforms is constructed as candidate signal templates. This list of potential waveforms is called the \textit{template bank}.
The specific number of required templates and specific locations of each template within the parameter space are the subject of much study in both  \ac{CBC}~\cite{1996PhRvD..53.3033B,1996PhRvD..53.6749O,1999PhRvD..60b2002O,1998PhRvD..57.2101B,Allen:2005fk,Harry:2009ea} and continuous \ac{GW} fields~\cite{1998PhRvD..57.2101B,2007PhRvD..75b3004P,2009PhRvD..79j4017M}. A template is considered a \textit{matched template} if it produces a $\rho$ greater than some set threshold $\rho_{\text{thr}}$ at any point in the given time series data.
The computational cost of calculating $\rho$ and comparing the value to $\rho_{\text{thr}}$ for all $M$ time steps for a template bank of $N$ templates is $O(NM\log M)$.

%If we consider $N$ templates in the frequency domain,the \ac{FFT} need only be applied to the data and to revert $\rho(f)$ to the time domain $\rho(t)$, which reduces the complexity from $O(M^2)$ to $O(M\log N_{\text{th}})$~\cite{cooley1967historical}. The calculation of $\rho$ itself has a complexity of $O(M)$. This corresponds to the $k_2$ steps in the previous discussion. Consequently, the final complexity for the whole process is $O(N N_{\text{th}}\log N_{\text{th}})$. According to our algorithm, this means that in Step 2 (line~\ref{alg:step2S}-\ref{alg:step2E}) of Algorithm.~\ref{alg:GroGate}, the cost would be $O(M\log M)$, still preserving the advantage of the \ac{FFT}. This results in the overall complexity of 
%\begin{equation}
%    \label{equ:FFTquantumcost}
%    O\left( \sqrt{N} \left( M\log M + \log N \right) \right),
%\end{equation}
%with a $\sqrt{N}$ speed up comparing with the fastest classical algorithm.

%%%%%%%%%%%%%%%%%%%%%%%%%%%%%%%%%%%%%%%%%%%%%%%%%%
%%%%%%%%%%%%%%%%%%%%%%%%%%%%%%%%%%%%%%%%%%%%%%%%%%
\subsection{Grover's Algorithm}\label{sec:grover}

The speed-up provided by Grover's algorithm is proved in an oracle model: the algorithm is given access to an oracle, which returns whether or not a given input is a good match, and in the quantum version it is assumed to allow queries in superposition. One way to achieve this is to assume that the database of interest is pre-loaded into \ac{qRAM}~\cite{giovannetti2008qRAM}. This can be efficiently queried, however there remain doubts about the experimental feasibility of qRAM, as well as whether the advantage over classical techniques persists once all resources needed are taken into account~\cite{aaronson2015,Preskill2018,ciliberto2018,tang2021quantum}. Further, for the problem considered here, the size of the database is prohibitively large, and thus we require an explicit construction of the oracle. There are therefore two requirements for a speed-up in a problem of interest: there must be no classical algorithm giving an improvement over a brute force search, and it must be possible to construct an oracle for the problem considered. Further, the oracle should be efficient, meaning that the computational cost of implementing the oracle must scale at most polylogarithmically in the number of entries in the database.

In this section and elsewhere in the paper, we use the asymptotic notation $O$ and $\Omega$ common in computing science to discuss the running time or number of gates required. The statement that $O(f(N,M))$ gates are required means that the asymptotic scaling of the number of gates required is \emph{upper} bounded by the function $f(N,M)$ of the parameters $N$, $M$ characterising the size of the input. Similarly, $\Omega(f(N,M))$ denotes a \emph{lower} bound in the asymptotic scaling. Where possible we also go beyond asymptotic scaling and give the exact number of operations needed for particular examples, to illustrate the potential speed-up over classical techniques.

Grover's algorithm, proposed by Lov Grover in 1996 \cite{grover1996fast}, is a quantum algorithm providing a polynomial speed-up for search problems compared to classical techniques. A search problem is one in which the aim is to identify one or more \textit{marked entries}, i.e., those satisfying a specified criteria, from within an unstructured database. For a database with $N$ entries and exactly one marked entry, it is necessary to check $N/2$ entries on average before finding the marked entry; thus the required search time for a classical algorithm is $O(N)$~\cite{barnett2009quantum}. Grover's algorithm finds a solution in $O(\sqrt{N})$ search time. It was later proved that this is asymptotically optimal; $\Omega(\sqrt{N})$ %\footnote{$O$-notation and $\Omega$-notation are both asymptotic notations, where they represent the upper and lower bound of the complexity of an algorithm respectively.}
queries are required for a quantum algorithm to succeed with high probability~\cite{bennett1997strengths}. Grover's algorithm is covered in several introductory quantum computing texts, e.g.,~\cite{barnett2009quantum, nielsen2011quantum,kaye2007introduction,rieffel2011} but for the purposes of clarity we use the remainder of this section to outline the algorithm.

%Grover's algorithm establishes a gap in query complexity between classical and quantum computers in an oracle model. That is, it assumes that we have access to an oracle, a ``black box" which computes a desired function, but not necessarily a description of the function itself. The query complexity is then given by the number of calls required to the oracle. Grover's algorithm is covered in several introductory quantum computing texts, e.g.,~\cite{barnett2009quantum, nielsen2011quantum,kaye2007introduction} but for the purposes of clarity we use the remainder of this section to outline the algorithm. In Sec.~\ref{sec:psuedocode} we address the problem of how to construct an oracle specifically for matched-filtering.

We begin with some very brief introductory remarks introducing basic concepts and terminology in quantum computing. The fundamental carrier of quantum information is the qubit, the analogy to the classical bit. Physically this is a quantum system with two orthogonal states, which we label $\ket{0}$ and $\ket{1}$, and which are known as computational basis states. A quantum register is made up of an array of qubits. Any classical bit string may be encoded into qubits by encoding in the computational basis, simply by preparing $\ket{0}$ for ``0" and $\ket{1}$ for ``1", known as digital encoding. Quantum gates are reversible, due to unitarity of quantum evolution, and any classical reversible logic operation can be directly implemented as a transformation of computational basis states. Note that reversibility is not a restriction, as any classical irreversible computational may be performed reversibly, most straight-forwardly by simply retaining copies of the input~\cite{barnett2009quantum, nielsen2011quantum,kaye2007introduction}. Finally it is worth stating explicitly that quantum algorithms generically are probabilistic, succeeding with high probability. This is also not a limitation, as the probability of success can be boosted close to one by a few repetitions of the algorithm. Some commonly used states and operations are defined in Appendix \ref{sec:quantumgates}.

Grover's algorithm establishes a gap in query complexity between classical and quantum computers in an oracle model. That is, it assumes access to an oracle, a ``black box" which computes a desired function, but not necessarily a description of the function itself. The query complexity is then given by the number of calls required to the oracle. To cast the search problem as an oracle problem, a function $f(x)$ is defined which takes the value $f(x)=1$ if and only if $x$ is a marked entry in the database, otherwise $f(x)=0$. In the quantum case, this is implemented by a quantum black box or oracle $U_f$ that acts as follows on computational basis states:
\begin{equation}
    \label{equ:Ufa}
    U_f: \ket{x}\otimes\ket{d}\longmapsto\ket{x}\ket{d \oplus f(x)},
\end{equation}
where $\otimes$ represents the tensor product and $\oplus$ is bitwise addition modulo 2. The first register is an input register; the state $\ket{x}$ represents the input $x$, stored as a classical bit-string in the computational basis. The second register is an output register; after application of $U_f$, the evaluation of the function is contained here, shifted by the initial bitstring $d$. The key difference in the quantum case is that the oracle may be queried in superposition, that is, the input register may be prepared in a superposition over all input states. Note that if the output register is prepared in the state $\ket{-}$ (see Eq.~\ref{equ:quantumstate+}), the operation given in Eq.~\ref{equ:Ufa} is equivalent to the following procedure, known as phase kickback, on the input register alone:
\begin{equation}
    \label{equ:Uf}
    U_f: \ket{x}\longmapsto(-1)^{f(x)}\ket{x}.
\end{equation}
Although in the actual algorithm presented later we will need the output register for the oracle, in the following discussion, we prefer to use Eq.~\ref{equ:Uf} for the oracle evaluation for simplicity. 

\begin{figure*}[ht]
  \subfloat[The input state, represented by the red line. %represents the input initial state $\ket{s}$ as a vector in a plane spanned by the desired match $\ket{s}$ and undesired match $\ket{w_\perp}$.
  \label{fig:1A}]{
	\begin{minipage}[c][1\width]{
	   0.3\textwidth}
	   \centering
	   \includegraphics[width=1\textwidth]{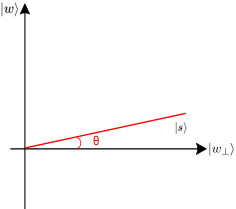}
	\end{minipage}}
 \hfill 	
  \subfloat[The state after the oracle is applied, represented by the blue line. %represents the result of the oracle acting on the initial state $\ket{s}$ which now is represented by the red dot dashed line.
  \label{fig:1B}]{
	\begin{minipage}[c][1\width]{
	   0.3\textwidth}
	   \centering
	   \includegraphics[width=1\textwidth]{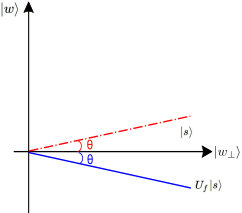}
	\end{minipage}}
 \hfill	
  \subfloat[The state after the diffusion operator, represented by the green line. %represents the result of the diffusion operator acting on the resulting state in B, $U_f\ket{s}$, which is now represented by the blue dotted line.
  \label{fig:1C}]{
	\begin{minipage}[c][1\width]{
	   0.3\textwidth}
	   \centering
	   \includegraphics[width=1\textwidth]{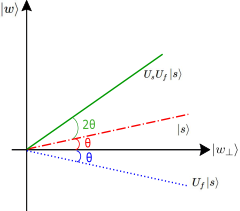}
	\end{minipage}}
\caption{We show how the input state $\ket{s}$ changes at different stages of Grover's algorithm. The two-dimensional space is spanned by the desired match $\ket{w}$ and undesired match $\ket{w_\perp}$. The solid lines represent the current state and the dotted lines represent the previous states.}
\end{figure*}

In the problem of searching in an unstructured database, the index of each entry in the database is represented as a computational basis state $\ket{i}$, and the input register is prepared in an equal superposition over all indices $\ket{s}$. Supposing that there are $N$ entries, the initial state of the input register can be expressed as:
\begin{equation}
    \label{equ:soo}
    \ket{s}=\frac{1}{\sqrt{N}}\sum^{N-1}_{i=0} \ket{i},
\end{equation}
where $1/\sqrt{N}$ represents the amplitude of each state in the superposition. This corresponds to an equal initial weighting of each entry. State $\ket{w}$ is used to represent an equal superposition of all the \emph{marked} entries in the database. In the following we will denote the number of marked entries by $r$%\footnote{We refer the reader to Table~\ref{tab:nomen} in the Appendix for a guide in the subscript notation used through the remaining sections.}
. The equal superposition of all the other entries of the database is denoted $\ket{w_\perp}$, which is perpendicular to the state $\ket{w}$. In terms of $\ket{w}$ and $\ket{w_\perp}$ the input state $\ket{s}$ may be rewritten as:
\begin{equation}
    \label{equ:som}
    \ket{s}=\sqrt{\frac{r}{N}}\ket{w}+\sqrt{\frac{N-r}{N}}\ket{w_{\perp }}.
\end{equation}
Now in order to increase the probability of finding one of the correct solutions, the next steps of Grover's algorithm are designed to increase the amplitude of the state $\ket{w}$ in the superposition. Throughout the algorithm the state of the input register remains within a real two-dimensional vector space spanned by $\ket{w}$ and $\ket{w_\perp}$. The initial state $\ket{s}$ is shown in Fig.~\ref{fig:1A}, where the angle between the states $\ket{w}$ and $\ket{s}$ is defined as
\begin{equation}
\label{equ:thetat}
   \theta=\arcsin\left(\braket{w}{s}\right) =\arcsin\left( {\sqrt{\frac{r}{N}}}\right). 
\end{equation}
After applying the oracle $U_f$, the input state $\ket{s}$ is transformed to
\begin{equation}
    \label{equ:ufs}
    U_f\ket{s}=-\sqrt{\frac{r}{N}}\ket{w}+\sqrt{\frac{N-r}{N}}\ket{w_{\perp }},
\end{equation}
which is equivalent to flipping the input state $\ket{s}$ with respect to the horizontal axis $\ket{w_\perp}$, as represented in Fig.~\ref{fig:1B}. This procedure itself however, does not make the desired state $\ket{w}$ more favourable in the measurement. Therefore, an additional diffusion unitary operator is applied as the second step, which is defined as
\begin{equation}
    \label{equ:Us}
    U_{s}=2\ket{s}\bra{s}-\hat{\rm I},
\end{equation}
where $\hat{\rm I}$ is the identity operator. Considering the state afterwards expressed in an orthonormal basis including the state $\ket{s}$, it is clear that this operator applies a minus sign to the amplitude of all states except $\ket{s}$. Analogously to the interpretation of the oracle, this is equivalent to reflecting the state of the register about the equal superposition state $\ket{s}$, as shown in Fig.~\ref{fig:1C}.
%\begin{equation}
 %   \label{equ:usufsp}
 %   U_{s}U_{f}\ket{s}=\eta_1\ket{s}-\eta_2\ket{s_\perp}.
%\end{equation}
%This shows the diffusion operator is equivalent to flipping the $U_f\ket{s}$ state with respect to the $\ket{s}$ state, as shown in Fig.~\ref{fig:1C}.

The overall effect of the Grover operator $\hat{G}$, defined as:
\begin{equation}
    \label{equ:Gdef}
    \hat{G}= U_{s}U_{f},
\end{equation}
is shown in Fig.~\ref{fig:1C}, and is equivalent to a rotation operator in the two-dimensional space spanned by $\ket{w}$ and $\ket{w_\perp}$:
\begin{equation}
    \label{equ:Gmatrix}
    \hat{G}=\begin{pmatrix}
\cos{2\theta} & -\sin{2\theta} \\
\sin{2\theta} & \cos{2\theta} 
\end{pmatrix}.
\end{equation}
After applying the Grover operator $k$ times, the input state would become
\begin{equation}
    \label{equ:Gks}
    \hat{G}^k \ket{s}= \sin{\big((2k+1)\theta\big)}\ket{w}+\cos{\big((2k+1)\theta\big)}\ket{w_\perp}
\end{equation}
and in order to maximise the probability of finding one of the desired matches comprising the superposition $\ket{w}$, the amplitude $\sin{\big((2k+1)\theta\big)}$ should be maximised. Thus the Grover operator is applied $k$ times such that $(2k+1)\theta\approx\ \pi/2$. This means that if the number $r$ of matching templates is known, for large values of $N/r$:
\begin{equation}
\label{equ:kt}
    k\approx\frac{\pi}{4}\sqrt{\frac{N}{r}} - \frac{1}{2}.
\end{equation}
After $k$ applications of Grover's algorithm, as all matching templates are in superposition, a measurement of the input register will return only one of them at random. To obtain additional matching templates the algorithm must be repeated $r\log r$ times~\cite{flajolet1992birthday}. 

%%%%%%%%%%%%%%%%%%%%%%%%%%%%%%%%%%%%%%%%%%%%%%%%
%%%%%%%%%%%%%%%%%%%%%%%%%%%%%%%%%%%%%%%%%%%%%%%%
\subsection{Quantum Counting}\label{sec:quantc}
In many cases the number of marked entries, $r$, is not known in advance. In this case there exist variants of Grover's algorithm which return a marked entry with $O\left(\sqrt{N/r}\right)$ applications of the oracle~\cite{brassard1998quantum,mosca2001counting}. The most relevant for our purposes is quantum counting, which uses a well-known primitive in quantum computing, quantum phase estimation~\cite{Kitaev_1997}, to estimate the eigenvalues $\pm 2 \theta$ of the Grover operator introduced in Eq.~\ref{equ:Gmatrix}. This in turn allows an estimate of $r$, and of the number of applications of the Grover operator needed to find a solution with high probability. $O(\sqrt{N})$ Grover iterations are sufficient to determine $r$ to an accuracy $O(\sqrt{r})$ with high probability. We complete this background section with an outline of quantum counting, and refer the reader again to texts ~\cite{barnett2009quantum, nielsen2011quantum,kaye2007introduction,rieffel2011} for more information.

%For a quantum counting process with $m+\log(2+1/2\epsilon)$ qubits, it determines Grover's operator's eigenvalue to an accuracy of $2^{-m}$ with probability $1-\epsilon$~\cite{nielsen2011quantum}. 

%There exists an alternative to the quantum counting algorithm that eliminates the extra counting register by incrementing the number of application of Grover's operator after failure to retrieve a solution. The search time of this algorithm is $O(\frac{1}{\theta})$, where $\theta$ denotes the eigenvalue of Grover's operator~\cite{kaye2007introduction}. Although this version of quantum counting has less space requirement, it needs an exhaustive search to identify a no solution situation. Because in gravitational wave research, the main question is if a solution exists, the rest of this paper will be based on the original version of quantum counting algorithm.

Recall that the Grover operator $\hat{G}$ acts as a rotation in the two-dimensional space spanned by $\ket{w}$ and $\ket{w_{\perp}}$, as given in Eq.~\ref{equ:Gmatrix}. The eigenvectors of $\hat{G}$ are 
\begin{equation}
  \label{equ:Geigenvector}  
    \ket{s_+}=\begin{pmatrix}
\frac{i}{\sqrt{2}} \\
\frac{1}{\sqrt{2}}
\end{pmatrix}, \qquad
\ket{s_-}=\begin{pmatrix}
\frac{-i}{\sqrt{2}} \\
\frac{1}{\sqrt{2}}
\end{pmatrix},
\end{equation}
with eigenvalues of $e^{2i\theta}$ and $e^{-2i\theta}$ respectively, and the input state in Eq.~(\ref{equ:soo}) may be written as an equal superposition of the two eigenstates, $\ket{s_+}$ and $\ket{s_-}$:
\begin{equation}
\label{equ:superst}
     \ket{s}=\frac{1}{\sqrt{2}}\left(\ket{s_+}+\ket{s_-}\right).
\end{equation}
%\smc{Working on notation: using $\ket{j}$ for the computational basis states ($\ket{i}$ used earlier, but $i$ is needed for the complex number $i$). The other summation index is now $l$, to keep $b$ for the actual result of measurement.}
Given an estimate of $\theta$, an estimate of the number of matching templates can be obtained through Eq.~\ref{equ:thetat}. Therefore, the problem of finding the number of desired templates is transformed into an eigenvalue estimation problem, which can be solved using quantum phase estimation~\cite{kaye2007introduction}. Phase estimation makes use of the quantum Fourier transform, which transforms between the computational basis $\{ \ket{j} \}$ and the Fourier basis, $\{ \ket{\Tilde{j}} \}$ defined as:
\begin{equation}
   \label{equ:QFT}
    \ket{\Tilde{j}}=\hat{U}_{QFT}\ket{j}= \sum_{l=0}^{2^p-1} \exp{\left(i\frac{2\pi jl}{2^p}\right)}\ket{l}.
\end{equation}
where $\hat{U}_{\text{QFT}}$ denotes the \ac{QFT}~\cite{nielsen2011quantum}.

\begin{figure}[htb]
{\includegraphics[width=\columnwidth]{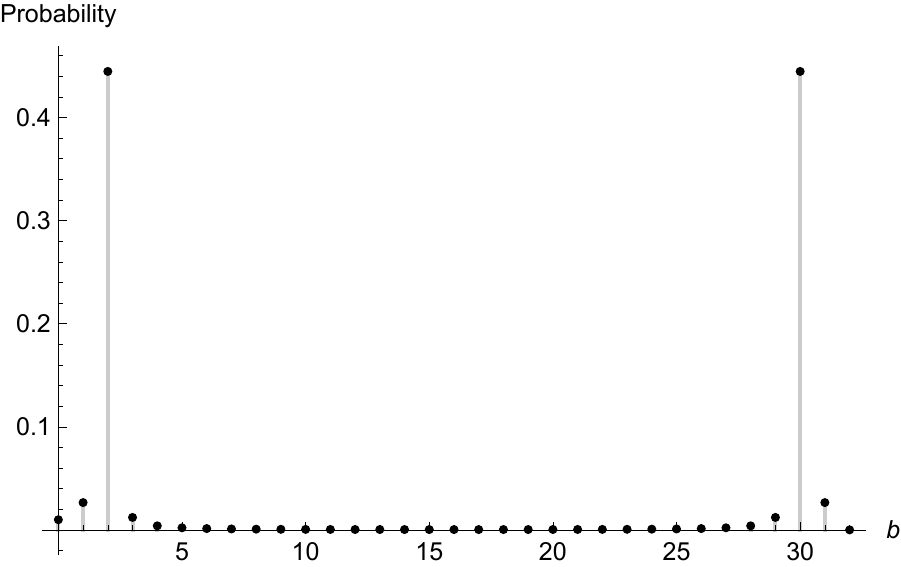}}
\caption{The probability distribution for each output value in the final measurement on a $5$-qubit counting register, with two matching entries in a $64$-entry database. The two peaks correspond to the two eigenstates defined in Eq.~\ref{equ:Geigenvector}. Constructive interference only happens for values close to $2^p \theta/\pi$ or $2^p(\pi -\theta)/\pi$, with destructive interference occuring elsewhere, resulting in this probability distribution.}\label{fig:distribution}
\end{figure}

In quantum counting an additional register, which we refer to as the \textit{counting register}, is needed to store the estimate of $\theta$. We denote the number of qubits in the register by $p$, which we leave unspecified for now. The counting register is first initialised in an equal superposition over all possible computational basis states: 
\begin{equation}
    \label{equ:H0N}
    \hat{H}^{\otimes p}\ket{0}^{\otimes p}=\frac{1}{2^{\frac{p}{2}}}\left(\ket{0}+\ket{1})\otimes...\otimes(\ket{0}+\ket{1}\right)=\sum_{j=0}^{2^p-1}\ket{j}.
\end{equation}
%A phase is naturally represented in the Fourier basis $\{\ket{\Tilde{a}}\}$, and is defined as follows:
%\begin{equation}
%   \label{equ:QFT}
%    \ket{\Tilde{a}}=\hat{U}_{QFT}\ket{a}= \sum_{b=0}^{2^p-1} \exp{\left(i\frac{2\pi ab}{2^p}\right)}\ket{b}.
%\end{equation}
%where $\hat{U}_{\text{QFT}}$ is the quantum Fourier transform \ac{QFT}~\cite{nielsen2011quantum}. Each basis element $\ket{\Tilde{a}}$ corresponds to a different phase $2 \pi a/2^p$ appearing in the superposition, so the inverse Fourier transform gives a way to extract information encoded as a phase.
Following this, Grover's operator is applied iteratively to the input state as before, where now the number of applications of the Grover gate is controlled by the counting register:
%\begin{widetext}
\begin{equation}
\begin{split}
    \label{equ:CG}
    &\sum_{j=0}^{2^p-1}C\text{-}\hat{G}^j\ket{j}\otimes \ket{s}
    \\=&\frac{1}{\sqrt{2}}\left(\sum_{j=0}^{2^p-1}e^{i2\theta j}\ket{j}\otimes\ket{s_+}+\sum_{j=0}^{2^p-1}e^{-i2\theta j}\ket{j}\otimes\ket{s_-}\right), \end{split}
\end{equation}
%\end{widetext}
where $C\text{-}\hat{G}^j$ represents applying the controlled Grover's operator $j$ times, giving: 
\begin{equation}
\begin{split}
    \label{equ:InverQFT}
   &\hat{U}^{-1}_{QFT}\sum_{j=0}^{2^p-1}C\text{-}\hat{G}^j\ket{j}\otimes \ket{s}\\
   =&\frac{1}{2^{p+\frac{1}{2}}}\sum_{j=0}^{2^p-1}\sum_{l=0}^{2^p-1} \left( e^{i2\pi j\left(\frac{\theta}{\pi}-\frac{l}{2^p}\right)}\ket{l}\otimes\ket{s_+}+e^{i2\pi j\left(\frac{\pi-\theta}{\pi}-\frac{l}{2^p}\right)}\ket{l}\otimes\ket{s_-}\right).
\end{split}
\end{equation}
A measurement of the counting register in the computational basis returns an integer value between $0$ and $2^p-1$, from which we can now extract the desired estimate of the phase. Intuitively, constructive interference occurs for those elements $\{\ket{l^\prime}\}$ for which
\begin{equation}
    \label{equ:bkbrelation}
    \frac{\theta}{\pi}-\frac{l^\prime}{2^p} \simeq 0,\qquad \text{or} \qquad\frac{\pi -\theta}{\pi}-\frac{l^\prime}{2^p} \simeq 0.
\end{equation}
We will only be interested in cases in which $r \ll N$, and thus $\theta \ll 1$. Therefore, the observed measurement outcome, which we denote $b$, gives an unambiguous estimate of $\theta$, denoted $\theta_\ast$ as follows: 
\begin{equation}
    \label{equ:thetadef}
    \theta_\ast =
\left\{
	\begin{array}{ll}
		\frac{b \pi}{2^p}, &  b \leq 2^{p-1} \\
    \pi - \frac{b \pi}{2^p}, & b > 2^{p-1}.
	\end{array}
\right.
\end{equation}
In reality, values of $b$ which differ slightly from the constructive interference condition are possible; an example of the probability distribution over $b$ is shown in Fig.~\ref{fig:distribution}. However, it may be shown that the measured value $b$ gives an estimate of $\theta$ to $m$ bits of accuracy with a probability of success at least $1-\epsilon$ if $p$ is chosen such that $p = m+\log(2+1/2\epsilon)$~\cite{nielsen2011quantum}. In quantum counting, an estimate of accuracy at least $O(N^{-1/2})$ is required, as $\theta$ itself is of this magnitude. Thus $m$ and $p$ are each of size $1/2 \log N$. The maximum number of applications of $\hat{G}$ is given by $2^p$, which is therefore $O(\sqrt{N})$. From the estimate of $\theta$ it is then possible to estimate $r$ and $k$, the number of applications of $\hat{G}$ needed to subsequently retrieve a marked entry with high probability. In the following sections we will discuss the choice of $p$ in more detail for the application to quantum matched filtering, going beyond the asymptotic analysis.

%%%%%%%%%%%%%%%%%%%%%%%%%%%%%%%%%%%%%%%%%%%%%%%%%%%%%%%%%%%%%%%%%%%%%%%%%%%%%%%%%%%
%%%%%%%%%%%%%%%%%%%%%%%%%%%%%%%%%%%%%%%%%%%%%%%%%%%%%%%%%%%%%%%%%%%%%%%%%%%%%%%%%%%%%%
\section{Quantum matched filtering algorithm}\label{sec:psuedocode}
In the previous section we introduced matched filtering, Grover's algorithm and its extension to quantum counting, and outlined the computational speed-up promised by quantum algorithms for the process of search in an unstructured database. In this section we argue that matched filtering for gravitational wave detection provides a natural application of quantum counting. We detail the pseudo-code of a possible implementation and prove that we can effectively construct the required oracle. We will also compare the computational cost of the quantum approach with the classical cost, taking account of the cost of the oracle evaluation, to evaluate overall complexity in each case and the relative speed-up.

As discussed in the previous section, matched filtering involves comparing data (originally) in the form of a time series against templates drawn from a template bank, searching for one or more matches above a pre-determined threshold. The templates for gravitational wave data analysis are well modelled by general relativity, and rather than performing comparisons against a previously populated database, these are calculated as part of the matched filtering procedure. Indeed the number of templates can be so large that pre-calculating and storing these in a database may have prohibitive memory requirements even in the classical case. Thus a pre-loaded database is not necessary for a quantum implementation, avoiding the need for a large amount of data to be loaded into \ac{qRAM}. Further, the steps needed in order to construct an oracle which determines whether or not a given template is a match are already part of the classical data analysis, and including these explicitly does not diminish the speed-up of the quantum approach, which we outline below.

We note that the cost of an oracle call (i.e., a single \ac{SNR} calculation) is not negligible; this scales with the observing time period and the frequency bandwidth over which the data is analysed, and must be taken into account in a full complexity analysis. Grover's algorithm does not speed up this step, and one might wonder whether a more sophisticated approach could give a speed up here also. We return to this in the discussion, and compare our quantum counting based approach to related tasks from the literature. What quantum counting \emph{can} do is improve the dependence of the overall computational cost on the number of templates, making previously intractable searches possible. In particular, as it is the spacing of templates, and therefore the overall number of templates required, that determines the sensitivity of the search, a quantum implementation of matched filtering based on quantum counting promises to enable the detection of signals too weak to detect by classical data processing techniques.

\subsection{Oracle construction}\label{sec:qmf}
We propose two applications of quantum counting to gravitational wave matched-filtering: one to determine whether there is a match at all, which is often the problem of interest in gravitational wave matched filtering; and the other to retrieve a matching template in the case in which there is at least one match. In order to apply quantum counting in each case, we first require an oracle to perform matched filtering with a predefined threshold. Thus we begin by detailing in Algorithm~\ref{alg:GroGate} the pseudo code to construct the Grover's gate. 

We begin with some preliminaries: recall that the number of templates is denoted by $N$, and the number of data points in the time-series by $M$. We choose a digital encoding, i.e. to represent the data and templates as classical bits encoded in the computational basis. Standard techniques exist to convert any, in general, irreversible classical logic circuit to a reversible one, which may readily be implemented on a quantum computer by replacing classical reversible gates by their quantum equivalents~\cite{bennett1997strengths, rieffel2011}. In general some scratch space is needed to aid in performing all calculations reversibly. We outline a specific implementation, making use of four registers: one \textit{data register} which must be of size (number of qubits) linear in $M$, and one \textit{index register}, which requires $\log N$ qubits. For intermediate calculations we specify also one register to hold the computed template, which must be of size linear in $M$, and one to hold the computed $\ac{SNR}$ value, which does not scale with $N$ or $M$ and is $O(1)$. We discuss the space requirements further in Section \ref{sec:discussion}.

The basic element of Grover's algorithm is a search over an index into a database, and an oracle construction must calculate the template from the index $i$, proceed to calculate the \ac{SNR}, and finally perform the check against the threshold value. We denote the number of gates needed to compute a template waveform from its parameters by $k_1$\footnote{We also need to specify the mapping from index to template parameters. For reasons of clarity we have not included this step explicitly here, but note that efficient algorithms exist (see~\cite{2007CQGra..24S.481P}), which add a modest complexity $O({\rm polylog}N)$. We discuss template placing in the example in Section~\ref{sec:GWSD} and \ref{sec:cw_example}.}. As each template consists of $M$ data points, this takes time linear in $M$. The number of gates needed to calculate the \ac{SNR} between a template and the data is denoted $k_2$. From the introduction in Sec.~\ref{sec:mf}, this requires time $O(M\log M)$. Finally, checking whether the result is above a given threshold $\rho_{\text{thr}}$, as defined in Sec.~\ref{sec:mf} takes $O(1)$ gates, and is denoted $k_3$. In this way, to compute the match against all templates we need $N \cdot (k_1+k_2+k_3)$ steps, which is the total classical cost. Consequently, the total computational complexity of the classical algorithm is $O(N M \log M)$.

To construct a quantum algorithm we require all the same steps, but in addition we need to erase the intermediate calculations, in order to disentangle the index register from everything else to complete the oracle application. The pseudo code for Grover's gate is given in Algorithm~\ref{alg:GroGate}.

%%%%%%%%%%%%%%%%%%%%%%%%%%%%%%%%%%%%%%%%%%%%%%%%%%%%%%%%%%%%
\begin{algorithm}[htb]
\caption{Grover's Gate \newline Complexity: $O(M\log M+\log N)$}\label{alg:GroGate}
\begin{algorithmic}[1]
\Function{Grover's Search algorithm}{$N$, $\ket{D}$, $\rho_{\textrm{thr}}$}
\label{GroverPseudo}
\Procedure{Oracle Construction}{}
\State \emph{Creating templates}:\label{alg:step1S}
\ForAll{$i<N$}
    \State{$ \ket{i}\ket{0} \gets \ket{i}\ket{T_i}$}
\EndFor\label{alg:step1E}
\State \emph{Comparison with the data}:\label{alg:step2S}
    \State$\ket{i}\ket{D}\ket{T_i}\ket{0} \gets \ket{i}\ket{D}\ket{T_i}\ket{\rho(i)}$
        \If {$\rho(i) <  \rho_{\textrm{thr}}$}
            \State $f(i)=0$
        \Else 
            \State $f(i)=1$
       %\State \textbf{close};
        \EndIf
 $\ket{i}\ket{D}\ket{T_i}\ket{\rho(i)} \gets (-1)^{f(i)}\ket{i}\ket{D}\ket{T_i}\ket{\rho(i)} $ \label{alg:step2E}
\State \emph{Dis-entangling registers}:\label{alg:step3S}
    \State $(-1)^{f(i)}\ket{i}\ket{D}\ket{T_i}\ket{\rho(i)} \gets (-1)^{f(i)}\ket{i}\ket{D}\ket{T_i}\ket{0}$
    \State $(-1)^{f(i)}\ket{i}\ket{D}\ket{T_i}\ket{0}\gets (-1)^{f(i)}\ket{i}\ket{D}\ket{0}\ket{0}$
\EndProcedure\label{alg:step3E}
\Procedure{Diffusion Operator}{}\label{alg:step4S}
\State $ \sum(-1)^{f(i)}\ket{i}  \gets \sum(2\ket{i}\bra{i}-\hat{\rm I})(-1)^{f(i)}\ket{i}$
\EndProcedure\label{alg:step4E}
\EndFunction
\end{algorithmic}
\end{algorithm}
%%%%%%%%%%%%%%%%%%%%%%%%%%%%%%%%%%%%%%%%%%%%%%%%%
\textbf{Discussion:} The following is the explanation for each step and the related computational cost for Algorithm~\ref{alg:GroGate}.

\textit{Oracle construction:} 
     \begin{itemize}
     \item Step 0: Initialisation 
     \newline [Cost: $O(M+\log N)$]
     \newline The initial state is comprised of four registers:
        \begin{equation}
            \label{step0}
            |\psi_{0}\rangle = \frac{1}{\sqrt{N}}\sum^N_i\ket{i}_I\ket{0}_{T}\ket{D}_D\ket{0}_{\rho},
        \end{equation}
        where the subscripts $I$, $T$, $D$ and $\rho$ represent the indices, templates, data, and the \ac{SNR} register respectively. Loading the data takes time linear in $M$, while initialising the index register to an equal superposition requires $O(\log N)$ gates \cite{nielsen2011quantum}.
         \item Step 1 (line~\ref{alg:step1S}-\ref{alg:step1E}): Creating templates
         \newline [Cost: $O(M)$]
               \newline Calculating the templates from the index is performed in superposition over all index values, at a cost of $k_1 \sim O(M)$ gates. The state after this step would be:
                \begin{equation}
                \label{step1}
                |\psi_{1}\rangle = \frac{1}{\sqrt{N}}\sum^N_i\ket{i}_I\ket{T_i}_T\ket{D}_D\ket{0}_{\rho}.
                \end{equation}
        \item Step 2 (line~\ref{alg:step2S}-\ref{alg:step2E}): Comparison with the data
        \newline [Cost: $O(M\log M)$]
             \newline The cost of calculating \ac{SNR} between the template and the data is $k_2 \sim O(M \log M)$. Finally we compare this result to a predetermined threshold to determine the value of $f(i)$; the function that determines whether a given template is a match or not at a cost of $k_3 \sim O(1)$. After this step the state becomes: 
            \begin{equation}
                \label{step2}
                |\psi_{2}\rangle = \frac{1}{\sqrt{N}}\sum^N_i(-1)^{f(i)}\ket{i}_I\ket{T_i}_T\ket{D}_D\ket{\rho(i)}_{\rho}.
            \end{equation}
            \item Step 3 (line~\ref{alg:step3S}-\ref{alg:step3E}): Disentangling registers 
            \newline [Cost: $O(M\log M)$]
            \newline The diffusion operator part of Grover's gate must act on the index register alone. If the index register is entangled with any other register, it will not have the desired effect. Therefore, we need to erase the computation of $\rho(i)$ and $T_i$ to remove any correlation between these registers and the index register. The erasure process is the reverse of the generation process. Accordingly, another $k_1+k_2$ cost is generated. The state after this step is 
            \begin{equation}
            \label{step3}
            |\psi_{3}\rangle = \frac{1}{\sqrt{N}}\sum^N_i(-1)^{f_i}\ket{i}_I\ket{0}_T\ket{D}_D\ket{0}_{\rho}.
            \end{equation}
       \item  Step 4 (line~\ref{alg:step4S}-\ref{alg:step4E}): Applying the Diffusion Operator
       \newline [Cost: $O(\log N)$]
        \newline This step is unique to the quantum algorithm and requires $\mathcal{O}(\log N)$ quantum gates~\cite{barenco1995elementary}. 
\end{itemize}

 \textit{Total Cost:}
    The total cost for a single oracle call is therefore
    \begin{equation}
    \label{equ:totalcostOracle}
        O\left(M\log M + \log N\right).
    \end{equation}
\subsection{Signal detection}

Now that we have constructed the required oracle for quantum matched filtering, we can readily apply quantum counting to problems of relevance to gravitational wave data analysis. Our application will firstly focus on whether there is a signal existing in the data, a common example in matched filtering. Once it has been identified that a signal is present a full Bayesian parameter analysis to determine the properties of the source must be performed separately~\cite{2015PhRvD..91d2003V,2019ApJS..241...27A}. Quantum counting returns $r_\ast$, an estimate of the number of matches, and so is ideally suited to this task. 

In order to identify if there is a signal, we are interested in four conditional probabilities: a \textit{true negative}, the probability of correctly returning that there is no template with an SNR above the predetermined threshold when there is no such template existing in the template bank, $P(r_\ast=0|r=0)$; a \textit{false negative}, the probability of identifying that there is no match when indeed there is no template in the template bank with an SNR above the predetermined threshold, $P(r_{\ast}=0|r> 0)$; a \textit{true positive}, the probability of identifying that there are templates with a SNR above the predetermined threshold when there exists such templates in the template bank, $P(r_{\ast}>0|r>0)$; and a \textit{false alarm}, the probability of identifying that there are templates with a SNR above the predetermined threshold when there no such template exists it template bank, $P(r_{\ast}>0|r=0)$.

Recall that quantum counting returns an integer $b$, between $0$ and $2^p-1$, from which we can estimate $\theta$ and therefore $r$. If there are no matches, perfect constructive interference occurs for $b=0$ in Eq.~\ref{equ:InverQFT} and $b=0$ is returned with certainty. Thus identifying whether or not there is a signal present simply requires us to check whether $b=0$ or $b \neq 0$. There will be some probability of returning $b=0$ in cases where there are in fact one or more matches, resulting in a false negative output of the algorithm. This may be made exponentially small through a constant number of repetitions. The resulting pseudocode is detailed in Algorithm~\ref{alg:GroPseudo}. As discussed earlier $2^p$ is required to be $O(\sqrt{N})$ to give a sufficient accuracy to distinguish $\theta$ from zero. At the end of this subsection we discuss further the impact of the choice of $p$ on the probability of a false negative.

\begin{algorithm}[htb]
\caption{Signal Detection\newline Complexity: $O\left((M\log M + \log N)\cdot\sqrt{N}\right)$}\label{alg:GroPseudo}
\begin{algorithmic}[1]
\State $\textit{p} \gets$ number of \textit{precision digits}
\State $\textit{N} \gets$ number of \textit{templates}
\State $i \gets $index of \textit{ templates}
\State $\rho_{\textrm{thr}} \gets$ \textit{threshold}
\State $\ket{0} \gets$ \textit{Data}  $\ket{D}$ 
\Procedure{Quantum Counting}{$p$, $N$, $\ket{D}$, $\rho_{\textrm{thr}}$}
\label{pro:QC}
\State \emph{Creating the counting register }:\label{alg:2step1S}
\State{$ \ket{i}\gets \ket{0}^p\ket{i}$}
\State{$ \ket{0}^p\ket{i}\gets \frac{1}{2^{p/2}}(\ket{0}+\ket{1})^p\otimes\ket{i}$}
\label{alg:2step1E}
\State \emph{Controlled Grover' gate}:\label{alg:2step2S}
\ForAll{$j<2^p$}
   \State $a \gets j$
    \Repeat
        \State Algorithm~\ref{alg:GroGate} \Call{Grover's Gate}{$N$, $\ket{D}$, $\rho_{\textrm{thr}}$}, $a--$
    \Until{$a==0$}
\EndFor
\State$\frac{1}{2^{p/2}}(\ket{0}+\ket{1})^n\otimes\ket{i} \gets \frac{1}{2^{(p+1)/2}}\sum ( e^{2i\theta j}\ket{j}\otimes\ket{s_+}+ e^{-2i\theta j}\ket{j}\otimes\ket{s_-})$\label{alg:2step2E}
\State \emph{Inverse Quantum Fourier Transform}:\label{alg:2step3S}
    \State$\frac{1}{2^{(p+1)/2}}\sum ( e^{2i\theta j}\ket{j}\otimes\ket{s_+}+ e^{-2i\theta j}\ket{j}\otimes\ket{s_-}) \gets \frac{1}{2^{p+1/2}}\sum\sum( e^{i2\pi j(\frac{\theta}{\pi}-\frac{l}{2^p})}\ket{l}\otimes\ket{s_+}+ e^{i2\pi j(\frac{\pi-\theta}{\pi}-\frac{l}{2^p})}\ket{l}\otimes\ket{s_-})$\label{alg:2step3E}
\State \emph{Measurement ($b$)}:\label{alg:2step4S}
\If{$b=0$}
\State \Return `There is no match.'
\Else{  $ r_\ast\gets \textbf{Round}\left[N\sin\left(\frac{b}{2^p}\pi\right)^2\right]$}
\EndIf
\If{$r_\ast=0$ }
\State $r_\ast\gets 1$
\EndIf\label{alg:2step4E} 
\EndProcedure
\end{algorithmic}
\end{algorithm}

\textbf{Discussion:} The following is the explanation for each step and the related computational cost for Algorithm~\ref{alg:GroPseudo}.

\textit{Signal detection:}
\begin{itemize}
    \item Step 0: Initialisation 
    \newline [Cost: $O(M+\log N)$]
    \newline This is the same as the step 0 in Algorithm~\ref{GroverPseudo}.
\end{itemize}
     \hspace{0.7cm} \textit{Quantum counting:}
        \begin{itemize}
        \item Step 1 (line~\ref{alg:2step1S}-\ref{alg:2step1E}): Creating counting register
        \newline[Cost: $O(\frac{1}{2}\log N)$]  
        \newline This step involves applying a Hadamard gate to each qubit incuring a cost of $p$. 
        
        \item Step 2 (line~\ref{alg:2step2S}-\ref{alg:2step2E}): Controlled Grover's Gate
        \newline [Cost: $O((M\log M+\log N)\sqrt{N})$]   
        \newline The cost is given by the largest number of iterations of Grover's gate needed, $2^{p}-1$. 
        \item Step 3 (line~\ref{alg:2step3S}-\ref{alg:2step3E}): Inverse quantum Fourier transform 
        \newline[Cost: $O((\log N)^2)$~\cite{barnett2009quantum}]   
        
        \item Step 4 (line~\ref{alg:2step4S}-\ref{alg:2step4E}): Measurement 
        \newline [Cost: $O(\frac{1}{2}\log N)$]  
        \newline The cost of measurement is $1$ for each counting qubit. For the actual measurement we obtain a value $b$. According to Eq.~\ref{equ:thetadef}, we can calculate an estimate of the number of matching templates $r_{\ast}$ based on Eq.~\ref{equ:thetat}. When there is no matching template, the probability of $b$ being measured as $0$ is $1$. Therefore, any other observed value of $b$ resulting in zero matching templates can be disregarded and thus corresponds to an estimate of one matching template.
        %\item Step 5: Output. 
        %\newline Cost: O(1)
        %\newline  Because the measurement is probabilistic, we do not know which eigenvalue the measurement corresponds to. Therefore, we need to run it through both relations and pick the sensible result.
        %\item Step 7: Repeat Quantum phase estimation. \newline Cost: $l*((2k_1+2k_2+k_3+34(\log_2 N)-31)*2^{P}+P+P^2+1.5)$
        %\newline Because quantum phase estimation output the ideal number of repetition with a probability. In order to achieve a sufficiently convincing result we need to run the algorithm a number of times: which we set as $l$, that is irrelevant to the data and template sizes. Every time we store the $k(i)$ into an array and the number that appeared the most times would be the $k$ we shall use in the search step. The process of finding the modal number in the array $k(i)$ is recorded as $l/2$ times as an average search time.
    \end{itemize}  
    \hspace{0.7cm} \textit{Total Cost:}
    \begin{equation}
    \label{equ:totalcost}
        O\left((M\log M + \log N)\cdot\sqrt{N}\right),
    \end{equation}
     
 We conclude by discussing the effect of the choice of $p$ on the probability of a false negative, denoted $\delta_{\text{n}}$.  According to the discussion in Sec.~\ref{sec:quantc}, $p$ can be written as
\begin{equation}
    \label{equ:prelation}
    2^p=c\sqrt{N},
\end{equation}
and the following discussion is on the choice of the constant $c$ and its effect on the probability of a false negative. We will use well-known bounds from the literature to motivate a particular choice of $c$, and therefore $p$. This is not a unique choice, but rather is a convenient one for which we can readily bound $\delta_n$.

In order to avoid triggering a false negative, the outcome of measurement of the counting register $b$ should not be $0$. According to~\cite{brassard1998quantum}, if $\tilde{b}$ is defined as either $\theta2^p/\pi$ or $(2^p-\theta2^p/\pi)$ (note that this is not in general an integer value), then the measured value $b$ differs from $\tilde{b}$ by $|b-\tilde{b}|\leq 1$ with a probability at least $8 /\pi^2$. Therefore, choosing $p$ such that $\tilde{b}-1>0$ ensures that the probability of a false negative is at most $1-8/\pi^2$. With this choice, Eq.~\ref{equ:thetat} and~\ref{equ:bkbrelation} thus gives the following restriction on $p$:
\begin{equation}
    \label{equ:pchoice1}
    2^p>\pi\sqrt{\frac{N}{r}}.
\end{equation}
This restriction is most stringent when $r=1$. Therefore, we obtain a lower bound for the choice of number of counting qubits:
\begin{equation}
  \label{equ:pchoiceF}
2^p>\pi\sqrt{N}.
\end{equation}

With this choice of $p$ we can obtain a slightly tighter bound on the false negative probability as follows. From Eq.~\ref{equ:falneg1}, the probability of a false negative when there exists one or more templates can be expressed as:
\begin{equation}
\begin{aligned}
  \label{equ:falsenegative1}
    \delta_{\text{n}}=P(b=0|r> 0)&=\frac{1}{2^{2p}}\frac{N\sin^2(2^p\theta)}{r}
    \leq \frac{1}{2^{2p}}\frac{N}{r}.
\end{aligned}
\end{equation}
With the choice in Eq.~\ref{equ:pchoiceF}, this probability is inversely proportional to $r$, and for all $r$ is bounded by:
\begin{equation}
  \label{equ:falsenegative2}
    \delta_{\text{n}}<\frac{1}{\pi^2}.
\end{equation}

We conclude that the signal detection algorithm based on quantum counting has a false alarm probability of 0 under all conditions, and a false negative probability of $1/\pi^2$, given the condition in Eq.~\ref{equ:pchoiceF} is met.
%Given the condition in Eq.~\ref{equ:pchoiceF} is met, we conclude that the probabilities of a false alarm and a false negative occurring are 1 and $1/\pi^2$ respectively. 

 If the false negative rate is $\delta_{\text{n}}$ for each run, by repeating the whole procedure $\ell$ times, the probability of obtaining $b=0$ every time is $\delta_{\text{n}}^{\ell}$. Therefore, the total tolerance of our procedure would be $\delta_{\text{n}}^{\ell} < \pi^{-2\ell}$. With a repetition logarithmic to its tolerance, the total complexity of the procedure is $O(\ell \pi \sqrt{N})$. 

In gravitational wave research, practical applications normally involve between $10^4$ to $10^{12}$ templates~\cite{DalCanton:2017ala,2019ApJ...875..122A}. With the lower bound of the number of templates, $10^4$, $p$ can be chosen to be 9 according to Eq.~\ref{equ:pchoiceF}. In the classical case, the computational cost is approximately $10^4$ oracle evaluations, while in the quantum case, 512 evaluations suffice for a single run of the signal detection algorithm. There is therefore an order of magnitude difference in cost even for cases with the lowest number of templates. The upper most extreme case that has been analysed has $10^{12}$ templates, in which $p$ would be chosen as $22$, resulting in a computational cost of around $10^{7}$ oracle evaluations. As a specific example, for a false negative probability of $\pi^{-12} \simeq 10^{-6}$ (one in a million) a total of $6 \times 2^{22} \simeq 3 \times 10^7$ evaluations are required. To reduce this to a one in a billion chance of a false negative, $9$ repetitions of the algorithm are needed, or a total of around $4.5 \times 10^7$ oracle evaluations. This is orders of magnitude smaller than the classical cost of $10^{12}$.

%%%%%%%%%%%%%%%%%%%%%%%%%%%%%%%%%%%%%%%%%%%%%%%%%%%%%%%%%%%%%%%%%%%%%%%%%%%%%%%%%%%%%

\subsection{Retrieving matched templates}\label{sec:RMTAlg}
In the case of a successful signal detection (the identification of 1 or more matching templates), we might wish to further examine its corresponding parameters using (one of) the matching templates. In this section, we will provide a pseudo algorithm to retrieve one or all matching templates.

The procedure to retrieve matching templates is based on Grover's algorithm in Algorithm.~\ref{alg:GroGate} and the result $r_\ast$ of Algorithm~\ref{alg:GroPseudo}. This is not the only way to retrieve a matching template given an unknown number of matches \cite{mosca2001counting}, but we anticipate that for most applications the signal detection algorithm would run first in order to determine whether there is any match above threshold. In any potential subsequent attempt to retrieve a matching template it is then natural to use the estimate $r_\ast$ already obtained.

\begin{algorithm}[htb]
\caption{Template retrieval \newline Complexity: $O\left((M\log M + \log N)\cdot\sqrt{N}\right)$}
\label{alg:templateretreiving}
\begin{algorithmic}[1]
\State $\textit{N} \gets \textrm{number of }\textit{templates}$
\State $i \gets \textrm{index of} \textit{ templates}$
\State $\rho_{\textrm{thr}} \gets$ \textit{ threshold}
\State $\ket{0} \gets \textit{Data}$  $ \ket{D}$ 
\State $r_{\ast}\gets\textrm{number of }\textit{matched templates}$
\State \emph{Calculating the number of repetitions}:\label{alg:3step0S}
\State {$k_{\ast}\gets \textbf{Round}\left[\frac{\pi}{4}\sqrt{\frac{N}{r_{\ast}}}-\frac{1}{2}\right]$}\label{alg:3step0E}
\Procedure{Retrieve one template}{}\label{alg:3step1S}
    \Repeat
        \State Algorithm~\ref{alg:GroGate} \Call{Grover's Gate}{$N$, $\ket{D}$, $\rho_{\textrm{thr}}$}, $k_{\ast}--$
    \Until{$k_{\ast}==0$}
    \State \emph{Output}:
    \State $i_{\textrm{correct}}$\label{alg:3step1E}
\EndProcedure
\end{algorithmic}
\end{algorithm}

\textbf{Discussion:} The following is the explanation for each step and the related computational cost for Algorithm~\ref{alg:templateretreiving}.
\textit{Templates retrieval:}
\begin{itemize}
\item Step 0 (line~\ref{alg:3step0S}-\ref{alg:3step0E}): Calculating the number of repetitions
    \newline[Cost: $O\left(1\right)$]
    \newline The output $r_{\ast}$ from Algorithm~\ref{alg:GroPseudo} is imported into Algorithm~\ref{alg:templateretreiving}, and we then calculate the number of required repetitions of Algorithm~\ref{alg:GroGate} from Eq.~\ref{equ:kt}.
\item Procedure 1 (line~\ref{alg:3step1S}-\ref{alg:3step1E}): Retrieve one template 
    \newline [Cost: $O\left(\sqrt{N/r_{\ast}}\left(M\log M+\log N\right)\right)$]
    \newline Grover's algorithm, Algorithm~\ref{alg:GroGate}, will be repeated $k_{\ast}$ times to achieve the desired template index. The value of $k_{\ast}$ according to our previous discussion will be $O(\sqrt{N/r_{\ast}})$. 
    
    The total cost of Algorithm~\ref{alg:GroPseudo} and retrieving one template combined is:
    \begin{equation}
    \label{equ:totalcosti}
        O\left((M\log M + \log N)\cdot\sqrt{N}\right).
    \end{equation}
    \end{itemize}
    
    \begin{itemize}
    \item Procedure 2 %(line~\ref{alg:3step2S}-\ref{alg:3step2E})
    : Retrieve all matched templates 
    %\newline [Cost: $O\left(\log r_{\ast}\sqrt{Nr_{\ast}}\left(M\log M+\log N\right)\right)$]
    \newline  In the case where all the matched templates are required to be found, it is not as trivial as repeating Procedure~1 $r$ (assuming $r_{\ast}\approx r$) times because it samples with replacement. It is, instead, a coupon collector problem~\cite{flajolet1992birthday}, which requires $\Theta(r\log r)$ repetitions of Procedure~1. As long as the number of matching templates is small comparing with the total number of templates in the bank, the complexity is the same for both procedures.
\end{itemize}

We conclude this section by discussing the overall probability of failing to return a matched template following this procedure. Note that if this probability is less than $0.5$, then with a constant number of repetitions, it can be made negligibly small to ensure successful retrieval of a matched template\footnote{There is nothing special about $0.5$ here, as long as the probability of failure is bounded away from 1 this is enough; 0.5 is a convenient choice.}.

Without loss of generality we consider in the following analysis only one eigenvalue in Eq.~\ref{equ:InverQFT}, corresponding to $\ket{s_+}$. The corresponding probability distribution for different measured values $b$ is given in Appendix \ref{sec:Pt0}. In any given run of the procedure, the probability of returning a matched template according to Eq.~\ref{equ:Gks} is therefore given by:
\begin{equation}
    \label{equ:preturntemp}
    P(\textrm{Match}) = |\sin\left((2k_{\ast}+1)\theta\right)|^2,
\end{equation}
where $k_{\ast}$ is the number of Grover's applications calculated through Eq.~\ref{equ:kt} from the outcome $b$ of Algorithm~\ref{alg:GroPseudo} and corresponding estimates $\theta_\ast$, $r_\ast$. Using Eq.~\ref{equ:probabilityOG}, the overall probability of failing to retrieve a matched template is given by:
\begin{equation}
    \begin{aligned}
    \label{equ:failureretrieve}
      P(\text{Fail})&=\sum_{l=0}^{2^p}  P(\text{Fail}|b=l)P(b=l)\\
    &= \frac{1}{2^{2p}}\sum_{l=0}^{2^p} \bigg(\frac{\sin\Big(2^p \theta\Big)}{\sin (\theta-\frac{\pi l}{2^p})}\bigg)^2|\cos\left((2k_{l}+1)\theta\right)|^2,
    \end{aligned}
\end{equation}
where $k_l$ is the number of repetitions of Grover's algorithm when $b=l$.

\begin{figure}[htb]
{\includegraphics[width=\columnwidth]{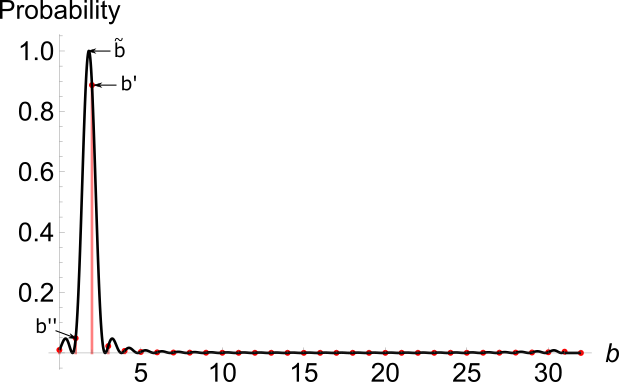}}
\caption{The red dotted line corresponds to the probability distribution for each state in a $5$-qubit counting register, with two templates matching in a $64$-template bank corresponding to one eigenvalue defined in Eq.~\ref{equ:Geigenvector}. The black line is plotted according to Eq.~\ref{equ:probabilityOG} as a continuous function. Each peak contains one $b$ state with a width of $1$, except for the central peak which has the two most probable $b$ states and a width of $2$. The upper integer $b$ state to $\tilde{b}$ is referred to as $b^\prime$ while the lower as $b^{\prime \prime}$. The curve peaks at either $2^p \theta/\pi$ or $2^p(\pi -\theta)/\pi$, depends on which eigenvalue the curve corresponds to, and it is labelled as $\tilde{b}$. }\label{fig:continousP}
\end{figure}

Let $b^\prime$ be the closest integer larger than $2^p \theta/\pi$, i.e. $b^\prime = {\lceil 2^p \theta/\pi \rceil} = 2^p \theta/\pi + \epsilon$ where $0 \leq \epsilon \leq 1$; and $b^{\prime \prime}$ the closest integer smaller than $2^p \theta/\pi$ such that $ b^{\prime \prime} =2^p \theta/\pi - (1 - \epsilon)$. $b^\prime$ and $b^{\prime \prime}$ are also the most probable values; recall that the probability that the measured $b$ value falls into the interval of $|b-\tilde{b}|\leq 1$ is larger than $8/\pi^2$~\cite{brassard1998quantum}. This is illustrated in Fig.~\ref{fig:continousP} based on Eq.~\ref{equ:probabilityOG} where the central peak contains the two most probable $b$ states.

Now an upper bound for $P(\text{Fail})$ is given by only considering the probability of successfully retrieving a template for these two most probable outcomes:
\begin{equation}
    \begin{aligned}
    \label{equ:failureretrieveUpBound0}
      P(\text{Fail})&< P(b^\prime) P(\text{Fail}|b^\prime)+P(b^{\prime \prime}) P(\text{Fail}|b^{\prime \prime})\\&+(1-P(b^\prime)-P(b^{\prime \prime})).
    \end{aligned}
\end{equation}
Now, to estimate $P(\text{Fail}|b^\prime)$, note using Eq.~\ref{equ:kt} that:
\begin{equation}
\begin{aligned}
\label{equ:kint}
k_{b^\prime}&=\left[\frac{\pi}{4\theta_{\ast}}-\frac{1}{2}\right]\\
&=\frac{\pi}{4\theta_{\ast}}-\frac{1}{2}\pm\epsilon_k, \\
&=\frac{2^{p-2}}{b^\prime}-\frac{1}{2}\pm\epsilon_k, \\
\end{aligned}
\end{equation}
where in the second line $0\leq\epsilon_k\leq0.5$, and in the third line we have used Eq.~\ref{equ:thetadef}. In the context of gravitational wave searches, i.e. $N\gg r$, the small angle approximation can be applied and consequently, $\theta\approx\sqrt{r/N}$. Thus
\begin{equation}
\begin{aligned}
(2 k_{b^\prime} + 1) \theta &= \frac{2^{p-1}}{b^\prime} \theta \pm 2\epsilon_k  \theta \\
&= \frac{\tilde{b}}{b^\prime} \frac{\pi}{2} + O \left( \sqrt{\frac{r}{N}} \right),
\end{aligned}
\end{equation}
from which we obtain using Eq.~\ref{equ:preturntemp} 
\begin{equation}
    \begin{aligned}
    \label{equ:bthfailureretrieve1}
      P(\text{Fail}|b^\prime) %&=1-P(\text{Match}|b^\prime)\\
      &=1-|\sin\left((2k_{b^\prime}+1)\theta\right)|^2\\
      &=\left|\cos\left(\frac{\tilde{b}}{b^\prime}\frac{\pi}{2}\right)\right|^2 +O \left( \sqrt{\frac{r}{N}} \right) \\
      &=\left|\cos\left(\frac{b^\prime-\epsilon}{b^\prime}\frac{\pi}{2}\right)\right|^2 + O \left( \sqrt{\frac{r}{N}} \right)\\
      &=\left|\sin\left(\frac{\epsilon}{b^\prime}\frac{\pi}{2}\right)\right|^2 +O \left( \sqrt{\frac{r}{N}} \right).
    \end{aligned}
\end{equation}
We can also rewrite $P(b^\prime)$ as follows:
\begin{equation}
   \begin{aligned}
   P(b^\prime) &= \frac{1}{2^{2p}} \bigg(\frac{\sin\Big(2^p \theta\Big)}{\sin (\theta-\frac{\pi b^\prime}{2^p})}\bigg)^2 \\
   &= \frac{1}{2^{2p}} \left(\frac{\sin\Big(\tilde{b} \pi\Big)}{\sin \left(\frac{\pi}{2^p} \epsilon \right)}\right)^2 \\
   &\simeq \left(\frac{\sin\Big(\epsilon \pi \Big)}{\pi \epsilon}\right)^2
   \end{aligned}
\end{equation}
where in the last line we have used the small angle approximation for $\pi \epsilon/2^p$, and $\tilde{b} = b^\prime - \epsilon$. With similar arguments for $b^{\prime \prime}$, the bound becomes:
\begin{equation}
    \begin{aligned}
    \label{equ:failureretrieveUpBound}
      P(\text{Fail})<&1 - \left(\frac{\sin\Big( \pi\epsilon
      \Big)}{\pi\epsilon}\right)^2 \left( \cos\left(\frac{\epsilon}{b^\prime}\frac{\pi}{2}\right)\right)^2
      \\&-\left(\frac{\sin\Big(\pi (1-\epsilon)
      \Big)}{\pi(1-\epsilon)}\right)^2\left(\cos\left(\frac{1-\epsilon}{b^{\prime \prime}}\frac{\pi}{2}\right)\right)^2 + O \left( \sqrt{\frac{r}{N}} \right)
    \end{aligned}
\end{equation}

\begin{figure}
\includegraphics[width=0.5\textwidth]{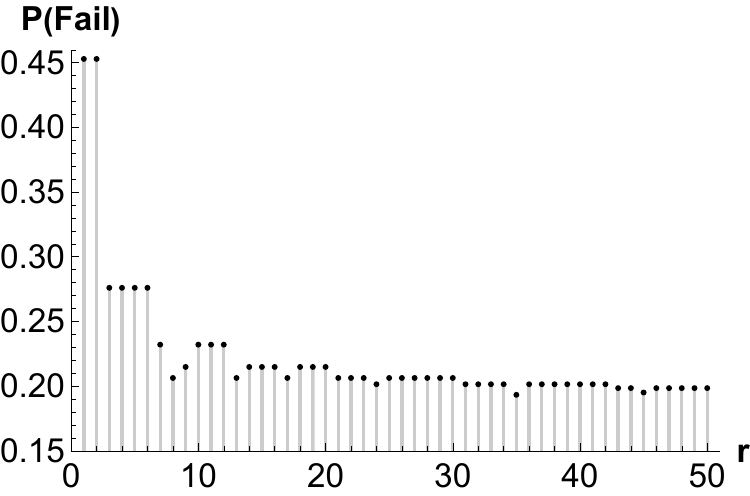}
\caption{This shows for large $N$, the joint probability of obtaining outcome $b$ and subsequently failing to retrieve a matched template is bounded by $0.45$ for different number of matching templates $r$.\label{fig:failprobGen} }
\end{figure}

Recall from Eq.~\ref{equ:pchoiceF}, we choose $p= \lceil \log_2(\pi \sqrt{N}) \rceil$. It is convenient to express this as $p=\log_2(\pi\sqrt{N})+\epsilon_p$, where $0<\epsilon_p<1$. Therefore $\tilde{b}$ may be written:
\begin{equation}
\begin{aligned}
\label{equ:btilda}
    \tilde{b}&=\frac{2^p\theta}{\pi}\\
    &=\frac{\pi \sqrt{N} 2^{\epsilon_p}}{\pi}\sqrt{\frac{r}{N}}\\
    &=2^{\epsilon_p}\sqrt{r}.
\end{aligned}
    \end{equation}
Recall that $b^\prime=\lceil{\tilde{b}}\rceil$, and so $b^\prime$, $\epsilon$ become:
\begin{equation}
    \label{equ:bprime}
    b^\prime=\lceil{2^{\epsilon_p}\sqrt{r}}\rceil; \quad \epsilon=\lceil2^{\epsilon_p}\sqrt{r}\rceil-2^{\epsilon_p}\sqrt{r}.
\end{equation}
Thus for each $r$ we can write Eq. \ref{equ:failureretrieveUpBound} in terms of a single parameter, $\epsilon_p$, between $0$ and $1$ (neglecting the $O(\sqrt{r/N})$ term). We optimise this numerically and plot the bound for various values of $r$ in Fig. \ref{fig:failprobGen}.
In all cases this is less than $0.453$, the value found numerically for $r=1$, ensuring the probability of successfully retrieving a template is no smaller than:
\begin{equation}
    \label{equ:Psuccess}
    P(\text{Success})\geq0.547 .
\end{equation}

Note that for large $r$ (but still requiring $r << N$),
\begin{equation*}
P(\text{Fail}|b^\prime) \simeq P(\text{Fail}|b^{\prime \prime}) \simeq \sin^2 \left(\frac{1}{\sqrt{r}} \frac{\pi}{2} \right) \simeq O \left( \frac{1}{r} \right)
\end{equation*}
and thus we can expect the bound on the probability of failure to decrease with $r$ to a limit given by:
\begin{equation}
\begin{aligned}
    P(\text{Fail}) &< 1 - P(b^\prime) - P(b^{\prime \prime}) + O \left( \frac{1}{r} \right) \\
    &= 1 - \frac{8}{\pi^2} + O \left( \frac{1}{r} \right).
\end{aligned}
\end{equation}
%\begin{equation}
%\begin{aligned}
%\label{equ:kint}
%k_{\tilde{b}}&=\left[\frac{\pi}{4\theta}-\frac{1}{2}\right]\\
%&=\frac{\pi}{4}\sqrt{\frac{N}{r}}-\frac{1}{2}\pm\epsilon_k,
%\end{aligned}
%\end{equation}
%where $0\leq\epsilon_p\leq0.5$. $\epsilon_p$ is negligible comparing with $\sqrt{N/r}$. Therefore, it is sufficient to use the relation that:
%\begin{equation}
%   \label{equ:kfinal}
%    (2k_{\tilde{b}}+1)\theta=\frac{\pi}{2}.
%\end{equation}
%According to Eq.~\ref{equ:thetadef}, $k_{\ast}$ is related to the measurement $b_{\ast}$ through:
%COnsequestly, the relationship between $k_{\ast}$ and the measured resulting $b_{\ast}$ can be obtained based on Eq.~\ref{equ:thetadef}:
%\begin{equation}
%\begin{aligned}
%\label{equ:inverserelation}
%(2k_{\ast}+1\pm\epsilon_k)\frac{b_{\ast} \pi}{2^p}&=\frac{\pi}{2}\\
% 2k_{\ast}+1&=\frac{2^{p-1}}{b_{\ast}}\mp \epsilon_k .
%\end{aligned}
%\end{equation}
%According to Eq.~\ref{equ:bprime} and Eq.~\ref{equ:pchoiceF}, $2^{p-1}/b^\prime$, $2^{p-1}/b^{\prime\prime}$ and $2^{p-1}/\tilde{b}$ are of $O(\sqrt{N/r})$. Therefore, $\epsilon_k$ is negligible. Based on this inverse relationship between $k_{b^\prime}l$ and $b^\prime$ in Eq.~\ref{equ:inverserelation}, we obtain:

\begin{figure}
\includegraphics[width=0.5\textwidth]{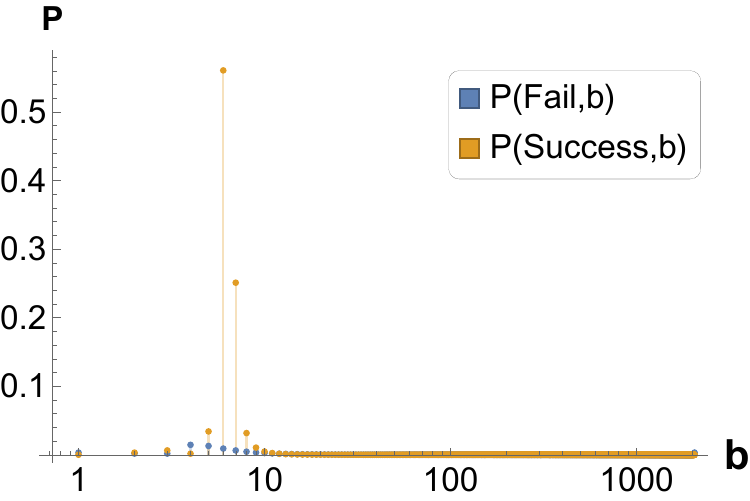}
\caption{For the case of a template bank with $2^{17}$ templates, and $r=9$, the joint probability of obtaining outcome $b$ and subsequently failing to or succeeding at retrieving a matched template are plotted in blue and yellow respectively. The total probability of $P(\text{Fail})\approx 0.34<0.5$.\label{fig:failprob} }
\end{figure}

We here provide a specific example of the total probability of failing to retrieve a matching template corresponding to Eq.~\ref{equ:failureretrieve} in Fig.~\ref{fig:failprob}. This example has a template bank of $2^{17}$ templates, with $r=9$, a real gravitational wave signal GW150914 that will be discussed in Sec.~\ref{sec:cbcexample}. The total failing probability $P(\text{Fail})\approx 0.34<0.5$. Therefore, with a constant number of repetitions of Alg.~\ref{alg:GroPseudo} and Alg.~\ref{alg:templateretreiving}, we are guaranteed with a matched template returned at a complexity of $O\left((M\log M + \log N)\cdot\sqrt{N}\right)$. This is less than the classical cost of $O\left(N M\log M\right)$. Therefore, we conclude that our quantum algorithm offers a $\sqrt{N}$ speed up with a practical oracle when the number of matching templates is small compared with the total number of templates in the bank.

%%%%%%%%%%%%%%%%%%%%%%%%%%%%%%%%%%%%%%%%%%%%%%%%
%%%%%%%%%%%%%%%%%%%%%%%%%%%%%%%%%%%%%%%%%%%%%%%%
\section{Example using Qiskit}\label{sec:qizkitexample}
%\begin{itemize}
%\item define model (small number of qubits and exact matching)
%\item one template matching
%\item multiple templates matching
%\item focus on results
%\item run on the actual quantum processor and compare the results, efficiency and gate costs
%\end{itemize}
In this section, we will present our proof of principle model of template matching on a quantum computer using IBM's Qiskit library \cite{Qiskit} and their quantum computer simulator \textit{ibmq\_qasm\_simulator}\footnote{The QasmSimulator backend is designed to mimic an actual device. It executes a Qiskit QuantumCircuit and returns a count dictionary containing the final values of any classical registers in the circuit.}. For the uninitiated reader, Appendix~\ref{sec:quantumgates} details relevant quantum computing fundamentals that are referred to throughout the following section. 

Matching to real gravitational wave data requires a much larger quantum processor than is currently available; in Section \ref{sec:cbcexample} we will present a classical simulation of matching to actual detector data using python. Later we also discuss the space requirements of the matched filtering algorithm. Here, in order to demonstrate the basic features of a realisation on a quantum processor, we implement a simplified algorithm in which we imagine the data is an $n$-bit string and the templates are all possible $n$-bit strings. This means that the templates themselves are identical to the index, and there is no need to explicitly perform the template generation steps (Algorithm~\ref{alg:GroGate} Step~$1$). We consider that a template is a match to the data if the bit strings are identical, however to simulate the possibility of non-exact matches, we disregard the $q$ lowest order bits and require only the $n-q$ highest order bits to match. The choice of $q$ is analogous to the choice of threshold SNR value $\rho_{\text{thr}}$ in the main algorithm. The proof of principle demonstration presented here is thus an example of string matching, a problem considered in \cite{ramesh2003string,montanaro2017quantum,niroula2021quantum}.

The data consists of an $n$-qubit string stored in binary form in the data register $\ket{D}$, where the first $q$ qubits are ignored allowing for $2^q$ matching templates among $2^n$ total templates. Hadamard gates are used to initialise the template register $\ket{T}$ to store a superposition of all possible $n$-bit templates. The output qubit $\ket{d}$ in Eq.~\ref{equ:Ufa} is stored in the ancilla register $\ket{A}$. An extra counting register with $p$ qubits is added for the quantum counting procedure. 

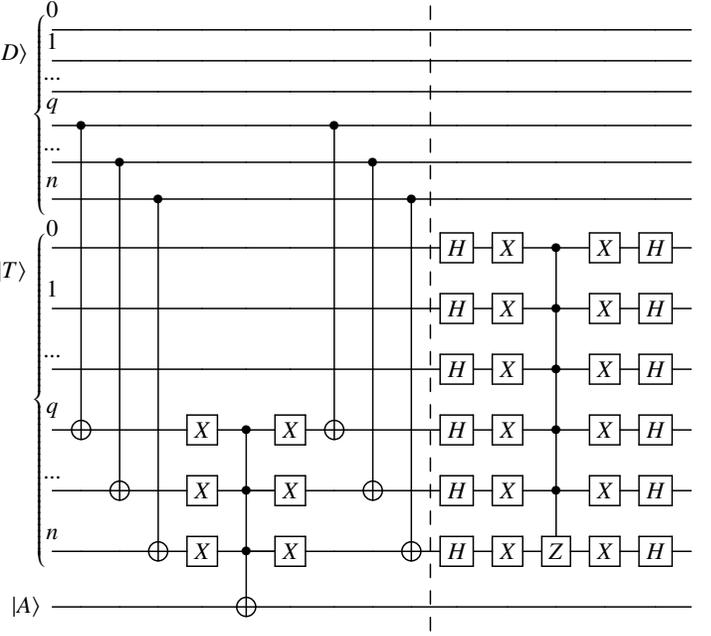
\begin{figure}[h!]
 \Qcircuit @C=0.8em @R=1.3em {
   \ustick{0} & \qw & \qw& \qw & \qw& \qw & \qw & \qw & \qw  \barrier[0.8em]{12} & \qw & \qw  & \qw& \qw & \qw& \qw & \qw\\
    \ustick{1} & \qw & \qw & \qw& \qw & \qw& \qw & \qw& \qw     & \qw& \qw & \qw& \qw & \qw& \qw & \qw\\
    \ustick{...} & \qw & \qw & \qw& \qw & \qw& \qw & \qw& \qw & \qw& \qw & \qw& \qw & \qw& \qw & \qw\\
    \ustick{q} & \ctrl{6} & \qw & \qw & \qw& \qw& \qw&\ctrl{6} & \qw & \qw& \qw & \qw& \qw & \qw& \qw & \qw\\
    \ustick{...} & \qw & \ctrl{6}& \qw & \qw& \qw& \qw& \qw& \ctrl{6}& \qw & \qw& \qw& \qw & \qw& \qw & \qw\\
    \ustick{n} & \qw & \qw & \ctrl{6}& \qw& \qw& \qw& \qw& \qw & \ctrl{6}& \qw& \qw& \qw & \qw& \qw & \qw
    \inputgroupv{1}{6}{1.1em}{1em}{\ket{D}}\\
    \ustick{0} & \qw & \qw& \qw & \qw& \qw& \qw & \qw & \qw& \qw&\gate{H}& \gate{X}&\ctrl{1}& \gate{X}&\gate{H}& \qw\\
    \ustick{1} & \qw & \qw& \qw & \qw& \qw& \qw & \qw & \qw& \qw&\gate{H}& \gate{X}&\ctrl{1}& \gate{X}&\gate{H}& \qw\\
    \ustick{...} & \qw & \qw & \qw& \qw& \qw& \qw & \qw & \qw& \qw&\gate{H}& \gate{X}&\ctrl{1}& \gate{X}&\gate{H}& \qw\\
    \ustick{q} & \targ & \qw & \qw& \gate{X}&\ctrl{1}& \gate{X}& \targ & \qw & \qw&\gate{H}& \gate{X}&\ctrl{1}& \gate{X}&\gate{H}& \qw\\
    \ustick{...} & \qw & \targ& \qw& \gate{X}&\ctrl{1}& \gate{X}\qw & \qw & \targ& \qw&\gate{H}& \gate{X}&\ctrl{1}& \gate{X}&\gate{H}& \qw\\
    \ustick{n} & \qw & \qw & \targ& \gate{X}&\ctrl{1}& \gate{X}\qw& \qw & \qw& \targ &\gate{H}& \gate{X}&\gate{Z}& \gate{X}&\gate{H}& \qw
    \inputgroupv{7}{12}{1.1em}{1em}{\ket{T}}\\
    \lstick{\ket{A}} & \qw & \qw& \qw & \qw &\targ&\qw& \qw& \qw& \qw & \qw& \qw& \qw& \qw & \qw& \qw 
  }
  \caption{Quantum circuit diagram for our multiple-template matching oracle and the diffusion operator, which are separated by the vertical dashed line. The $\ket{D}$ and $\ket{T}$ variables represent the data and template registers respectively and $\ket{A}$ is the ancilla qubit. The numbers label the $i^{\text{th}}$ qubit in the respective register. To simulate multiple matches, the oracle does not act on the first $q$ qubits. When there is only one matching template $q$ would be $0$.}
\label{fig:oracle circuit}
\end{figure}

In our template matching oracle, which is presented in Fig.~\ref{fig:oracle circuit}, we match the template register and the data register qubit-by-qubit using CNOT gates. In the case of an exact match, all the qubits in the template register would be turned into state $\ket{0}$. Therefore, after bit flipping, we can use a multiple-control-NOT gate to realise phase kickback on the ancillary qubit initialised into the $\ket{-}$ state. The diffusion operator is constructed by a combination of Hadamard gates, NOT gates and a $C^n$-$Z$ gate, and is illustrated in Fig.~\ref{fig:oracle circuit}. 
%A $Z$ gate is applied to the lowest qubit in the counting register to prevent a global phase of $\pi$ that would be problematic in the quantum counting stage \cm{can this problematic issue be explained in more detail?}.

In gravitational wave searches, the true signal parameters will lie somewhere within the template bank parameter space and no template will be identical to the signal. Therefore, a predetermined $\rho_{\text{thr}}$ is chosen as the threshold in Algorithm.~\ref{alg:GroGate}. The number of templates possessing $\rho$ over this threshold, if there are any, is unknown. Since the optimal number of applications of Grover's search algorithm is dependent on the number of templates with $\rho$ over the threshold, we need to apply the quantum counting algorithm first. 
%%%%%%%%%%%%%%%%%%%%%%%%%%%%%%%%%%%%%%%%%%%%%%%%%%%%%%%%%%%%%%%%%%%%%%%%%%%%%%%%%%%%%%
%%%%%%%%%%%%%%%%%%%%%%%%%%%%%%%%%%%%%%%%%%%%%%%%%%%%%%%%%%%%%%%%%%

To demonstrate a proof of principle of our algorithm, we implement this simplified version with a range of qubits for data and omission, allowing for multiple templates matching. For each pair $n$, $q$, we run the quantum counting algorithm first, in order to estimate the number of matches $r$, and then Grover's algorithm to find a match. From the output of the quantum counting algorithm, we take the most probable value of $b$ to calculate an estimated $r_\ast$ and $k_\ast$ for the template retrieval phase. For each algorithm the experiment is trialed 2048 times and the output of the simulator gives a set of probabilities calculated from the number of occurrences of each possible measured value. The results are presented in Table.~\ref{tab:simresult}. The number of counting qubits, $p$, is based on Eq.~(\ref{equ:pchoiceF}). When the number of qubits for the data, $n$, is small, $p$ is close to $n$. However, as $n$ increases, the difference between $n$ and $p$ increases as well, allowing us to maintain the speedup of $\sqrt{N}$ discussed in Sec.~\ref{sec:psuedocode}. The parameters $k_{\ast}$ and $k$ are the estimated and true number of applications of Grover's gate needed, given by the quantum counting process by Eq.~\ref{equ:thetadef} and Eq.~\ref{equ:kt} with $r=2^q$ respectively. The probability of the search process returning us with one of the matched templates given the most probable value of $b$ is over $78\%$ in all cases, and the estimated number of templates, $r_{\ast}$ differs from the actual number of matched templates, $2^q$, by no more than $2$.
%%%%%%%%%%%%%%%%%%%%%%%%%%%%%%%%%%%%%%%%%%%%%%%%%%%%%%%%%%%%%%%%%%%%%%%%%%%%%%%%%%%%%%%%%%%%%

\begin{table}[h!]
\resizebox{\columnwidth}{!}{%
\begin{tabular}{cccccccc}
\hline
\hline
ignored & data & counting & measured & Grover's & est. No. & Grover's &\\
qubits & length & qubits & count & iter. est. & templates & iter. theo. & $P$(\text{Succ.})\\
$q$ & $n$ & $p$ & $b$ & $k_{\ast}$ & $r_{\ast}$ & $k$ & \\	
\hline
 \multirow{5}{*}{0} 
&5 &	5&30 &	4&	1&	4&	0.9995\\
&6	&5&	1&	6&	1&	6&		0.9961\\
&7&	5	&1	&8&	1&	8&		0.9956\\
&8&	6	&	1	&12&	1&12&		1\\
&9	&7	&2&	17&	1&17&	0.9990\\
\hline
\multirow{6}{*}{1}   
&5 &	5&3 &	2&	3&	3&	0.9092
\\
&6	&5	&	30&	4	&2&	4	&	0.9985
\\
& 7	&6	&	61&	5&	3&	6	&	0.9619
\\
& 8	&6  &2  &8 &2	&8	&	0.9961
\\
&9&	7	&125&	10&	3&12	&	0.9365
\\
&10	&7	&126&	17&	2&17&	0.9995\\
\hline
\multirow{6}{*}{2} 
&5	&5	&	4&	1	&5&	2	&	0.7885
\\
&6	&5	&	29&	2	&5&	3	&	0.9072
\\
&7&	6&	60&	3&	5&	4&	0.8926

\\
&8&	6&	61&	5&	6&	6&	0.9688

\\
& 9&	7&	124&	7&	5&	8&	0.9429

\\
&10	&7	&125&	10&	6&12&	0.9395\\
\hline
\end{tabular}}
\caption{Trial runs of our algorithm with 2048 iterations on \textit{ibmq\_qasm\_simulator}. We compare the number of iterations Grover's algorithm should apply and the number of matched templates based on the measured result, to their theoretical counterparts across a range of data with different number of qubits with various number of omitted qubits in the matching process. We also state the $P$(Success) as the probability of our algorithm returning us with a matched template in the final search in each case. The number of counting qubits is the minimum allowed by Eq.~\ref{equ:pchoiceF} to minimise the false negative rate, $\delta_{\text{n}}$. \label{tab:simresult}}
\end{table}

\begin{figure}
\includegraphics[width=0.5\textwidth]{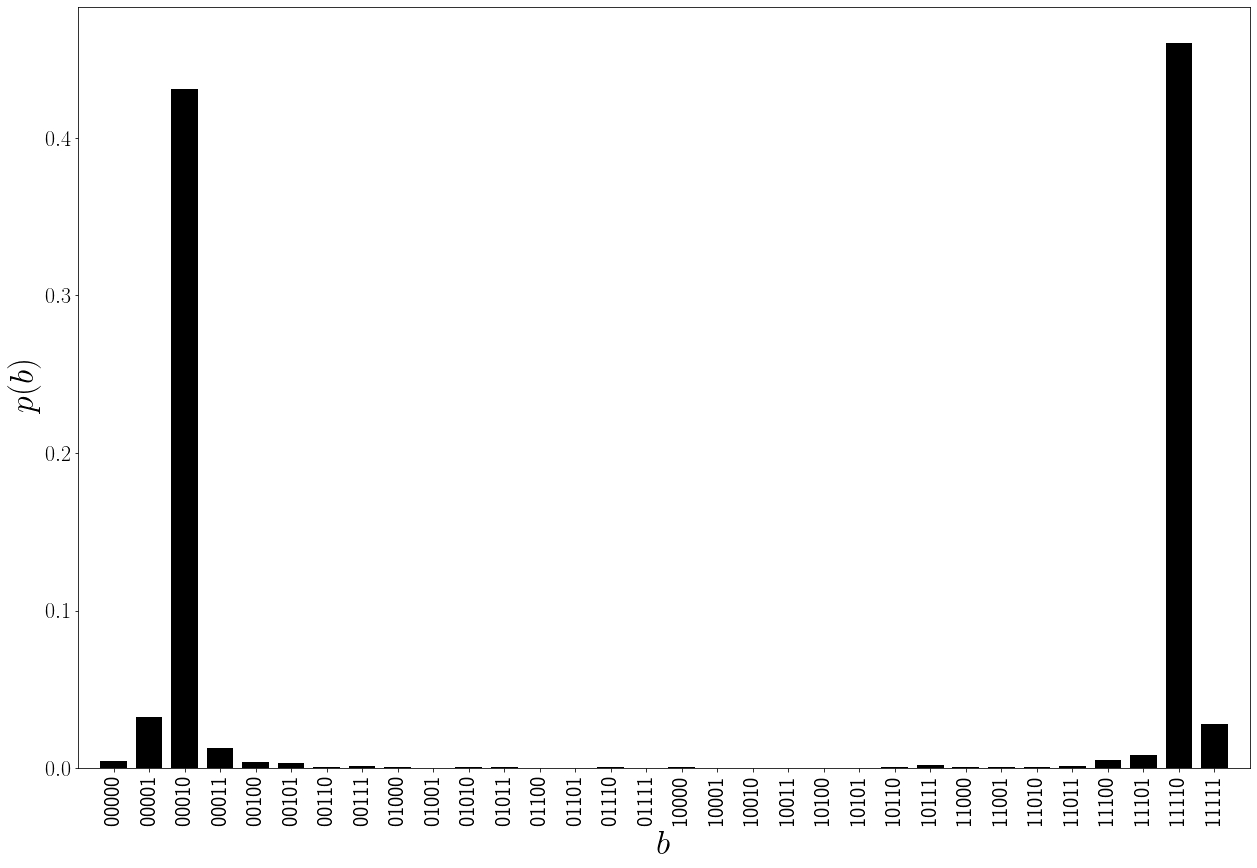}
\caption{The measurement of the quantum counting process for $6$-qubit data matching with a $5$-qubit counting register. The first qubit is ignored to allow for two templates matching. The theoretically most probable outcome $b$ in this case, according to Eq.~\ref{equ:bkbrelation}, should be either $2$ or $30$. The most probable measurement result is $11110$, which in decimal is 30.\label{fig:countingmeasure}}
 \end{figure}

\begin{figure}
\includegraphics[width=0.5\textwidth]{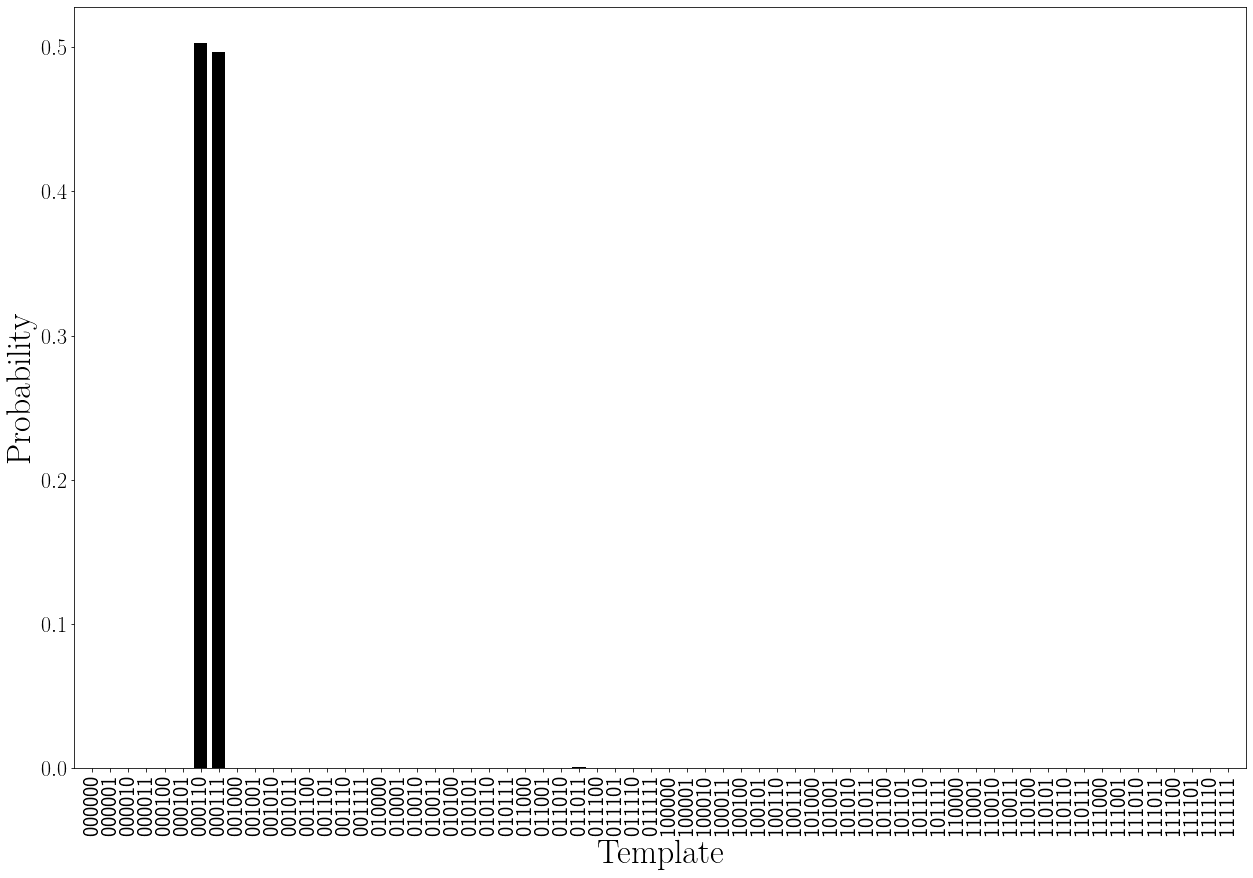}
\caption{The measurement of the Grover's search process for $6$-qubit data matching. The data is set as 000110 and the lowest qubit is ignored to allow for two templates matching. With $4$ iterations suggested by the quantum counting process as a numerical output, the two templates that meet the matching criteria are returned with a probability higher than $99\%$ altogether after $2048$ trials on \textit{ibmq\_qasm\_simulator}.\label{fig:grovermeansure}}
\end{figure}

A specific instance is illustrated in Fig.~\ref{fig:countingmeasure} and~\ref{fig:grovermeansure}. This case corresponds to $n=6$, $q=1$, and the data is fixed to be 000110. $q=1$ means that we look to find templates that match at least the last 5 qubits, i.e., 000110 and 000111. This is the same scenario as the analytical example we presented in Fig.~\ref{fig:distribution}, and described in Sec.~\ref{sec:quantc}. The result of the quantum counting process is shown in Fig.~\ref{fig:countingmeasure} where we can see that the measured values corresponding to the two eigenvalues from Eq.~\ref{equ:Geigenvector} are the most probable to be obtained. Converting the state indices from binary to decimal, our result is a bimodal distribution with $2$ modes: $2$ and $30$ are the locations of the mode peaks with a standard deviation less than $2$.  Both cases correspond to an estimate of $4$ for $k_{\ast}$, the same as the true value of $k$ calculated from the real number of templates.  Although this result does not exactly equal that given in Fig.~\ref{fig:distribution}, the fact that this algorithm is performed on a quantum simulator with limited number of runs needs to be taken into consideration.

In Fig.~\ref{fig:grovermeansure}, we show the result of the Grover's search process based on the result from Fig.~\ref{fig:countingmeasure}, in which the two matching templates are recovered with high probability in relation to other templates. Since they form an equal superposition, the two matched templates are assigned approximately equal probability. After performing 2048 trials of simulation in our results, the two matched templates constitute altogether a success probability $>99\%$.
%%%%%%%%%%%%%%%%%%%%%%%%%%%%%%%%%%%%%%%%%%%%%%%%%%%%%%%%%%%%%%%%%%%%%
 %%%%%%%%%%%%%%%%%%%%%%%%%%%%%%%%%%%%%%%%%%%%%%%%%%%%%%%%%%%%%%%%%%%%%%%%%%%%%%%%%%%%%%%%%%%%%%%%%%%%

%%%%%%%%%%%%%%%%%%%%%%%%%%%%%%%%%%%%%%%%%%%%%%%%
%%%%%%%%%%%%%%%%%%%%%%%%%%%%%%%%%%%%%%%%%%%%%%%%%
%\section{Example on Sine wave (python)}\label{sec:sineexample}
%
%\begin{itemize}
%\item continuous wave GW signal motivation
%\item why we don't just use the QFT? Our real CW signals (and CBCs) are not monochromatic
%\item Define sine wave model
%\item focus on showing results - number of steps taken
%\end{itemize}
%
%%%%%%%%%%%%%%%%%%%%%%%%%%%%%%%%%%%%%%%%%%%%%%%%
%%%%%%%%%%%%%%%%%%%%%%%%%%%%%%%%%%%%%%%%%%%%%%%%
\section{Example Search for GW150914}\label{sec:cbcexample}
\begin{figure}
	\centering
	\includegraphics[width=0.5\textwidth]{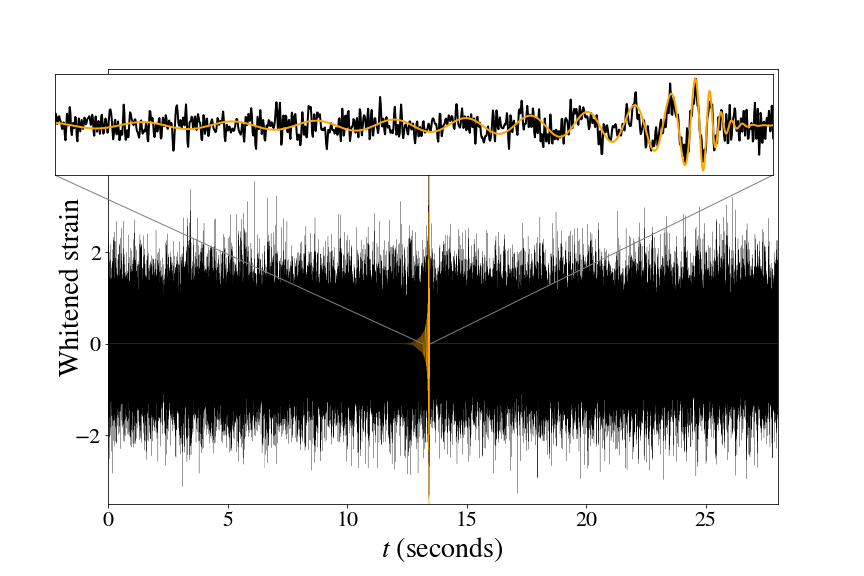}
	\caption{Whitened time-series data (black) of the gravitational wave event GW150914 sampled at $4096\,$Hz after a $20\,$Hz highpass filter overlaid by a signal template (orange) with component masses $m_{1}=35.6\,M_{\odot}$ and $m_{2}=30.6\,M_{\odot}$ and with zero aligned spin, taken from~\cite{abbott2019gwtc}. The signal can be more clearly seen in the $0.25\,$s plot in the upper panel.
	\label{fig:GW150914}}
\end{figure}

We now consider how this method can be used in the context of gravitational wave astronomy, namely the detection of the first gravitational wave event GW150914~\cite{2016PhRvL.116f1102A}. In this more complex scenario the data and template bank sizes are too large to be analyzed using IBM's Qiskit library, but we can compute the amplitudes of quantum states that correspond to the template and counting register at various stages of the algorithm described in Sec.~\ref{sec:psuedocode}. This is carried out on the \href{https://github.com/Fergus-Hayes/quantum-matched-filter}{\textit{quantum-matched-filter}} \textit{Python} code that is publicly available on Github.
The gravitational wave strain time-series data that we choose to analyse is from the LIGO Hanford detector and is centered around the GW150914 event time (GPS time 1126259462.4). It is $28$~s in duration and sampled at a rate of $4096\,$Hz. The data is initially whitened and passed through a high-pass filter with a $20\,$Hz lower cut-off frequency. The resulting time-series is shown in Fig.~\ref{fig:GW150914} in black. An approximate matching template is plotted overlaying the data in orange.
We perform our analysis on a bank of $2^{17}$ templates covering the 4-dimensional search space defined by the component masses $m_{1,2}$ and the aligned spin magnitudes $s_{1,2}$ of the binary system. We search these templates to find instances that correspond to matching templates.

We first consider applying the \textsc{Signal Detection} procedure of Alg.~\ref{alg:GroPseudo} to determine if a signal is present in the data, and acquire an estimate on the number of matching templates in Sec.~\ref{sec:GWSD}. 
In Sec.~\ref{sec:GWRMT} we show how to continue the analysis by using the \textsc{Template Retrieval} procedure of Alg.~\ref{alg:templateretreiving} to obtain matching templates.

\subsection{Signal Detection}\label{sec:GWSD}

\begin{figure}
	\centering
	\includegraphics[width=0.45\textwidth]{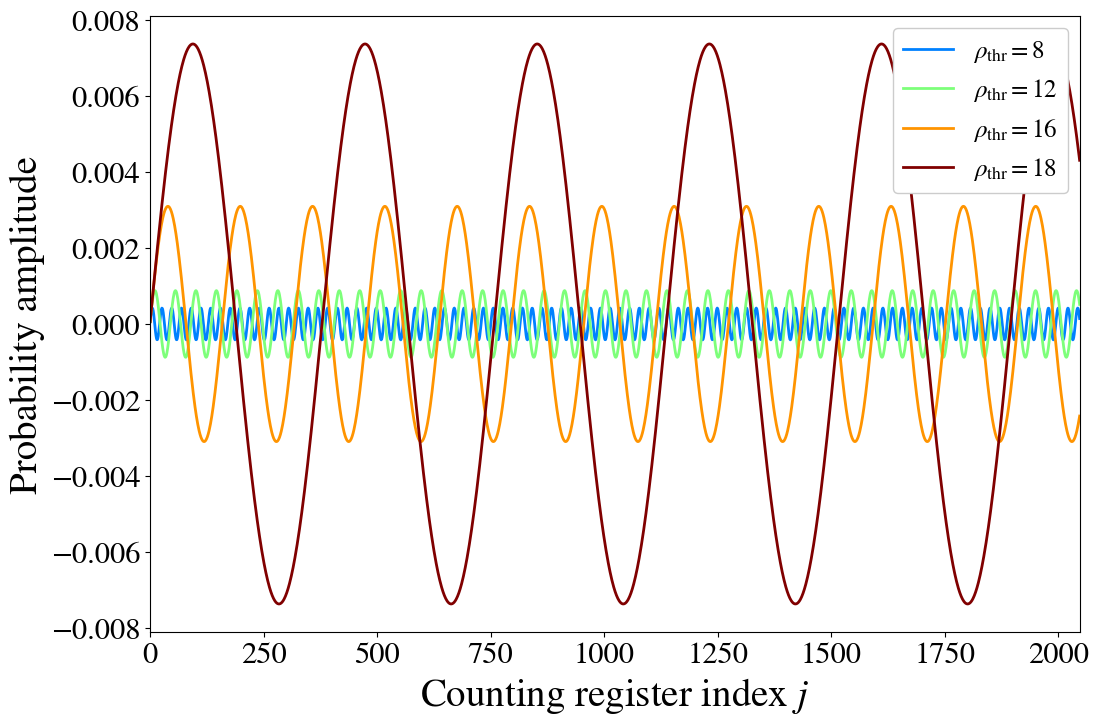}
	\caption{Probability amplitude of a single matching template over applications of the controlled Grover's gate specified in lines~\ref{alg:2step2S}-\ref{alg:2step2E} of Alg.~\ref{alg:GroPseudo} for the four instances of $\rho_{\text{thr}}=8,12,16,18$ given GW150914 data and with $p=11$. Larger $\rho_{\text{thr}}$ decreases the number of matching templates and therefore increases $k$. As all matching templates are amplified equally for each case, for a case with fewer matching templates the total amplitude is divided into fewer equal parts, leading to a larger amplitude for a matching template in comparison to cases with more matching templates.}
	\label{fig:psi1matches}
\end{figure}

First $|\psi_{0}\rangle$ from Eq.~\ref{step0} is initialised and the strain data is stored in $|D\rangle$.
The indices for each of the $N$ templates are represented by $|i\rangle$ and are put into superposition with the $2^p$ states in the counting register as described in Alg.~\ref{alg:GroPseudo} lines~\ref{alg:2step1S}-\ref{alg:2step1E}.
The controlled Grover's operator is applied to $|\psi_{0}\rangle$ as described by Alg.~\ref{alg:GroPseudo} lines~\ref{alg:2step2S}-\ref{alg:2step2E} to compare the templates to the data using Alg.~\ref{alg:GroGate} as a subroutine.
The templates are created from $|i\rangle$ to produce $|T_i\rangle$ as described in lines~\ref{alg:step1S}-\ref{alg:step1E} of Alg.~\ref{alg:GroGate}. 
Here this is done by using a look-up table that is computed prior to the analysis~\cite{canton2017designing} that accepts a given index as a key and returns the set of parameters $\{m_{1},m_{2},s_{1},s_{2}\}$ corresponding to the template. 
The parameters are then given to the phenomenological waveform model \textsc{IMRPhenomD} to produce the template~\cite{husa2016frequency,khan2016frequency,khan2019phenomenological}.
For a quantum computer implementation, we anticipate that this step would not be performed using such a look-up table, as this would rely on using qRAM.
Instead an algorithm is required that maps the $N$ template indices to their respective locations in the parameter space. 
The details of this algorithm are beyond the scope of this paper but it can be based on existing classical algorithms, such as those used for lattice-based template placement~\cite{prix2007template,cokelaer2007gravitational,2009PhRvD..79j4017M,brown2012detecting,harry2014investigating}, as any classical algorithm can be performed on a quantum computer and made reversible with at most polynomial overhead~\cite{bennett1997strengths}.

For each template in the bank, the oracle calculates $\rho$ for each time step using Eq.~\ref{equ:discretesnr} and applying the \ac{FFT} to produce $\{\rho_{i}(t_{1}),\dots,\rho_{i}(t_{M})\}$ where $M=28\times4096$ is the number of time steps.
A classical search algorithm is also written into the oracle to find $\rho^{\text{\text{\text{\text{max}}}}}_{i}=\max(\{\rho_{i}(t_{1}),\dots,\rho_{i}(t_{M})\})$.
We then simulate the phase kickback as described in Eq.~\ref{equ:Uf} giving $f(i)=1$ if template $i$ is a matching template, corresponding to $\rho^{\text{\text{\text{max}}}}_{i}\ge\rho_{\text{thr}}$ and $f(i)=0$ otherwise (non-matching template).
This can be written explicitly as:
\begin{equation}\label{equ:fsnr}
f(i) = 
\begin{cases}
    1\text{ if }\max\left(\frac{2}{M\Delta t}\left|\text{FFT}\left(\frac{\tilde{Q}_{c,i}(f)\tilde{h}(f)}{S_{n}(f))}\right)\right|\right)\ge \rho_{\text{thr}},\\
    0\text{ else.}
\end{cases}
\end{equation}
On a quantum computer, this function is evaluated for all templates in parallel, but is repeated $2^{p}-1$ times across the counting register.
The number of counting qubits is set to $p=11$, which is the fewest number of qubits in the counting register to meet the condition set in Eq.~\ref{equ:pchoice1}.
The probability amplitude of states that correspond to matching templates over each of these operations given GW150914 data is illustrated in Fig.~\ref{fig:psi1matches} for the analysis repeated with $\rho_{\text{thr}}=8,12,16,18$. Over successive iterations the probability amplitude of the states change according to Eq.~\ref{equ:Gks} with $\theta$ defined in Eq.~\ref{equ:thetat}.
With larger $\rho_{\text{thr}}$ there are fewer matching templates $r$ and the period of the probability amplitude's sinusoidal variation over the counting register states consequently increases. As all matching templates are amplified equally with each application, the probability amplitude is divided between fewer states with larger $\rho_{\text{thr}}$, which leads to the variations in the amplitude scale seen in Fig.~\ref{fig:psi1matches}.

\begin{figure}
	\centering
	\includegraphics[width=0.45\textwidth]{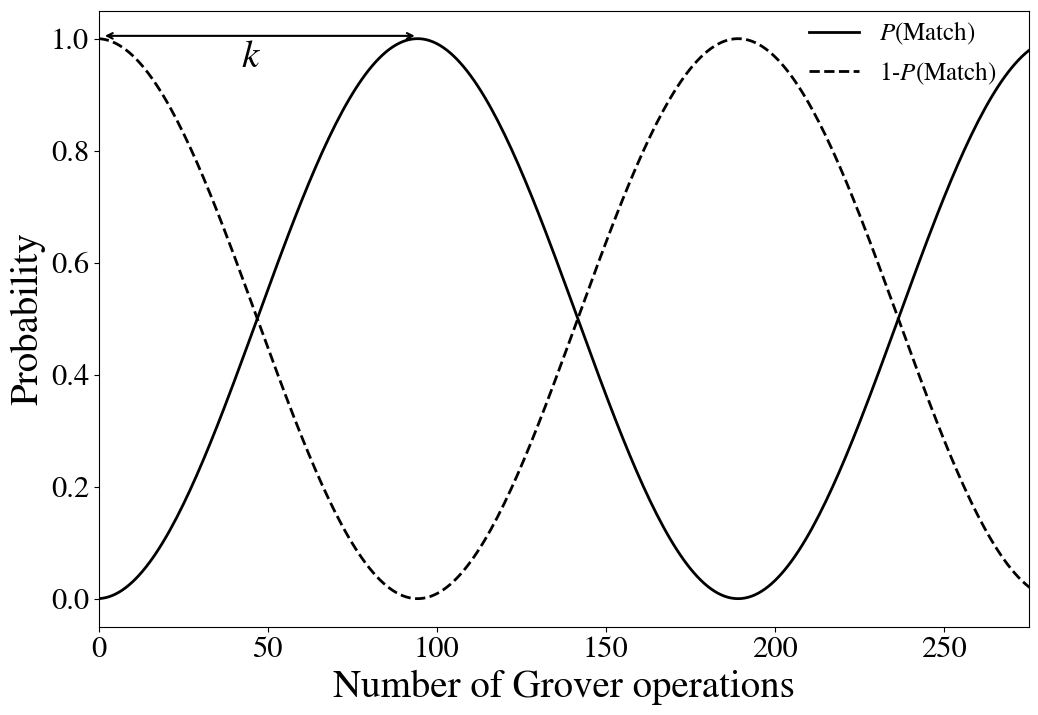}
	\caption{The probability of returning a matched template (solid line) and non-matching templates (dashed line) after the template register is measured after $k$ successive applications of Grover's operation given the case of $\rho_{\textrm{thr}}=18$ from Fig.~\ref{fig:psi1matches}. The probability amplitude of matching templates follows the sinusoid shown in Fig.~\ref{fig:psi1matches} for $\rho_{\textrm{thr}}=18$ while that of non-matching templates follow the same sinusoid but with a $\pi/2$ phase shift. The probability of returning a matching template is first maximised after $k$ Grover's operations.}
	\label{fig:psi1matchvnmatch}
\end{figure}

The amplitudes of the states that correspond to non-matching templates evolve in a similar sinusoidal fashion as the matching states as shown in Fig.~\ref{fig:psi1matches} but out of phase.
This is illustrated in Fig.~\ref{fig:psi1matchvnmatch} where the probability of recovering a matching template $P(\textrm{Match})$ (solid line) is compared to the probability of recovering a non-matching template (dashed line) over successive applications of Grover's operator for the case of $\rho_{\text{thr}}=18$.
Initially all states are equally probable so that the probability of returning a matching template is $r/N$, and evolve according to Eq.~\ref{equ:preturntemp} over Grover's operations.
The probability of returning a matching template is increased by applying Grover's operator successively until a maximum is reached after $k$ applications as defined in Eq.~\ref{equ:kt}.

\begin{figure}
	\centering
	\includegraphics[width=0.45\textwidth]{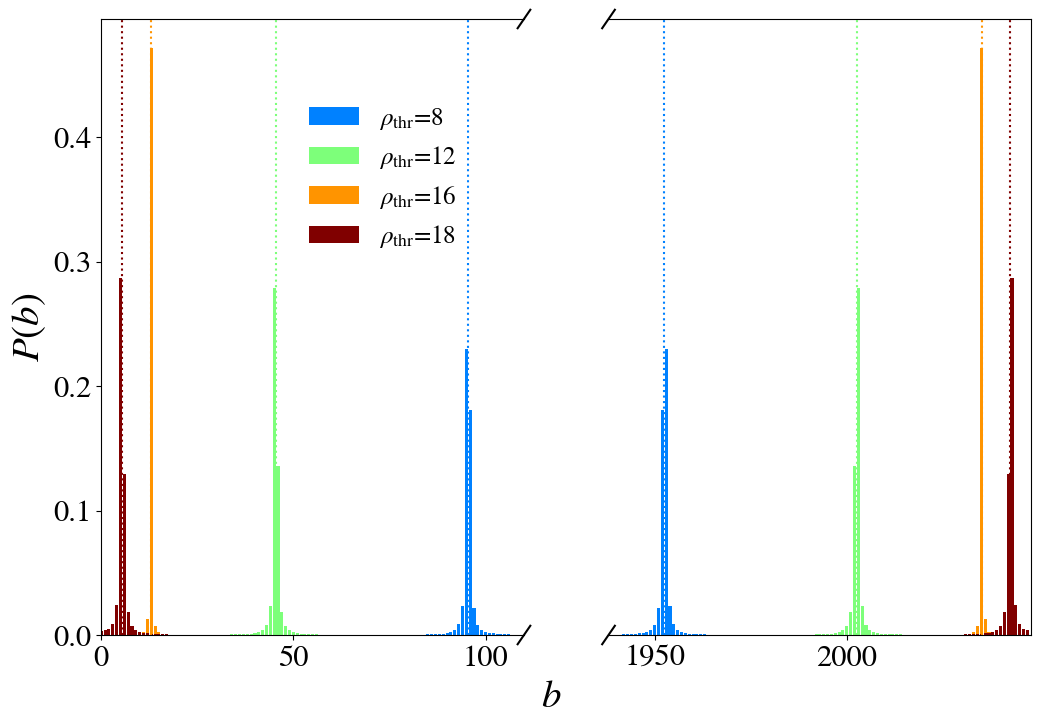}
	\caption{The probability of different outcomes $b$ of measuring the counting register after the inverse quantum Fourier transform is applied to the states in Fig.~\ref{fig:psi1matches}. This process is  described by lines~\ref{alg:2step3S}-\ref{alg:2step3E} for the different cases of $\rho_{\text{thr}}$ given $p=11$. The distributions are compared to the corresponding value of $\tilde{b}$ (dotted). The probability distributions corresponding to the two eigenvalues of Grover's operator are closer to $2^{p-1}$ for cases with more matched templates (lower $\rho_{\text{thr}}$). Cases with fewer matched templates are closer to the extremities of the range of $b$ and have an increased probability of not identifying any matched templates, corresponding to $P(b=0)$. This probability can be reduced by repeating the algorithm.}
	\label{fig:psi2b}
\end{figure}

\begin{figure}
	\centering
	\includegraphics[width=0.45\textwidth]{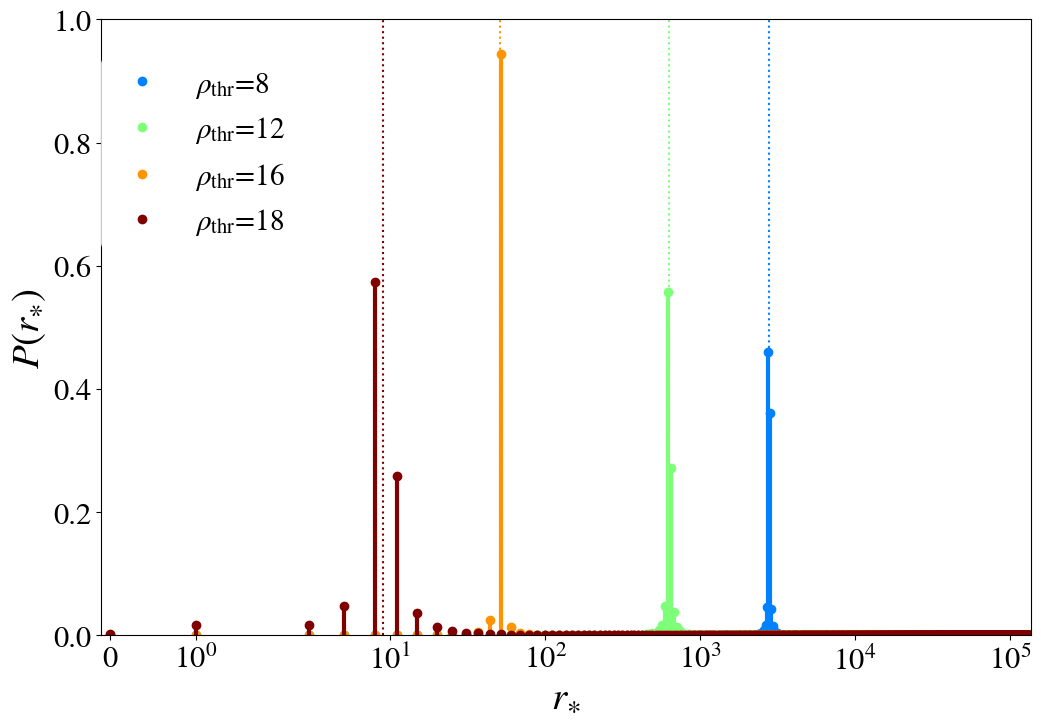}
	\caption{The probability distributions of outcomes from measuring the counting register from Fig.~\ref{fig:psi2b} transformed to estimates on the number of matching templates $r_{\ast}$ for each of the different cases of $\rho_{\text{thr}}$. The distributions are compared to the true number of matching templates $r$ (dotted).}
	\label{fig:psi2t}
\end{figure}

An estimate of the number of matching templates can be made from quantum counting as described in lines~\ref{alg:2step3S}-\ref{alg:2step3E} of Alg.~\ref{alg:GroPseudo} by applying the inverse QFT across the counting register states $\{|j\rangle\}$ to obtain $\{|l\rangle\}$.
Fig.~\ref{fig:psi2b} displays the probabilities of each outcome $b$ after a measurement is performed on the counting register for the different cases shown in Fig.~\ref{fig:psi1matches} with $p=11$.
The probability of different outcomes after measuring the counting register for the four different cases are compared to the non-integer value $\tilde{b}$, defined by the exact solutions of Eq.~\ref{equ:bkbrelation}, and plotted with a dotted line in Fig.~\ref{fig:psi2b}. The most probable outcome corresponds to $b'$ or $b''$ for each case, where the form of the distributions are governed by Eq.~\ref{equ:probabilityOG}. 
The outcome of measuring the counting register can equally be represented in terms of a prediction of the number of matching templates according to Eq.~\ref{equ:thetat} and Eq.~\ref{equ:bkbrelation} as shown in Fig.~\ref{fig:psi2t} for the example cases. For each $\rho_{\text{thr}}$ considered, the distributions peak near the actual number of matching templates. 
Notably, the probability of obtaining an outcome that corresponds to a non-zero number of matching templates is much greater than the probability of an outcome corresponding to zero matching templates for all cases. 
This is equivalent to the probability of obtaining an outcome other than $b=0$ in Fig.~\ref{fig:psi2b}. 
Obtaining an outcome of $b=0$ given the case where there are matching templates is a false negative,
the probability of which is governed by Eq.~\ref{equ:falsenegative2}.
Therefore the rate of false negatives (made in addition to that produced from the classical matched filtering approach) can be reduced by repeating the \textsc{Signal Detection} procedure. This should be compared to the case where there are no matching templates to identify. In this case the measurement of the counting register always results in $b=0$ corresponding to no matching templates. This negates the possibility of the analysis producing additional false alarms to the classical matched filtering approach as $P(r_{\ast}>0|r=0)=0$. 
If we only wish to determine if a signal is present in the data or not then the analysis can stop at this stage after the counting register is measured. The cost of determining this outcome requires $2^{p}-1$ enquiries of the oracle, in comparison to the $\sim O(N)$ calculations of $\{\rho(t_{1}),\dots,\rho(t_{M})\}$ from Eq.~\ref{equ:discretesnr} in the classical case.

%%%%%%%%%%%%%%%%%%%%%%%%%%%%%%%%%%%%%%%%%%%%%%%%%%%%%%%%%%%
\subsection{Retrieving Matching Templates}\label{sec:GWRMT}

\begin{figure}
	\centering
	\includegraphics[width=0.45\textwidth]{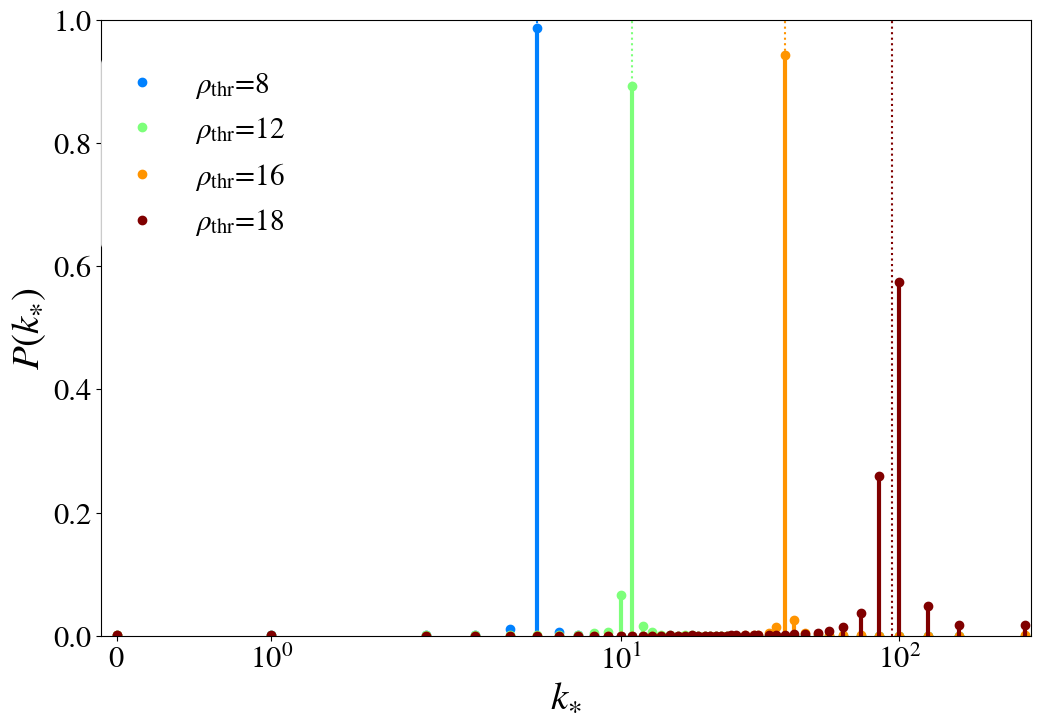}
	\caption{The probability distributions of outcomes from measuring the counting register from Fig.~\ref{fig:psi2b} transformed to estimates on the optimal number of Grover's applications $k_{\ast}$ for each of the different cases of $\rho_{\text{thr}}$. The probabilities are compared to the true $k$ (dotted) for each case.}
	\label{fig:psi2k}
\end{figure}

Similar to how the number of matching templates is estimated from the counting register's measurement outcome in Sec.~\ref{sec:GWSD}, the optimal number of Grover's operations is estimated using Eq.~\ref{equ:kt}. Fig.~\ref{fig:psi2k} shows the probability of obtaining different values of $k_{\ast}$ from the measurement for the same cases of $\rho_{\text{thr}}=8,12,16,18$ used in the previous section, and shows that the distributions peak around $k$, indicated by the dotted line. Fig.~\ref{fig:psi2k} is truncated at $(2^{p-1}-1)/2$, so as to exclude the outcome corresponding to zero matching templates and only consider outcomes of $b>0$.

Given the resulting $k_{\ast}$, the \textsc{Template Retrieval} procedure in Alg.~\ref{alg:templateretreiving} can be applied to obtain a matching template. This involves again initializing $|\psi_{0}\rangle$ from Eq.~\ref{step0} and applying \textsc{Grover's Gate} in Alg.~\ref{alg:GroGate} to this state iteratively $k_{\ast}$ times. This is done to maximize the probability that measuring the template register will return an index that corresponds to a matching template as illustrated in Fig.~\ref{fig:psi1matchvnmatch}. Each state that corresponds to a match will be amplified equally so that the probability of obtaining any given matching template is uniform.
For a given $k_{\ast}$, the probability of obtaining a matching template is governed by Eq.~\ref{equ:preturntemp}.

\begin{figure*}%[ht]
\centering
	\includegraphics[width=\textwidth]{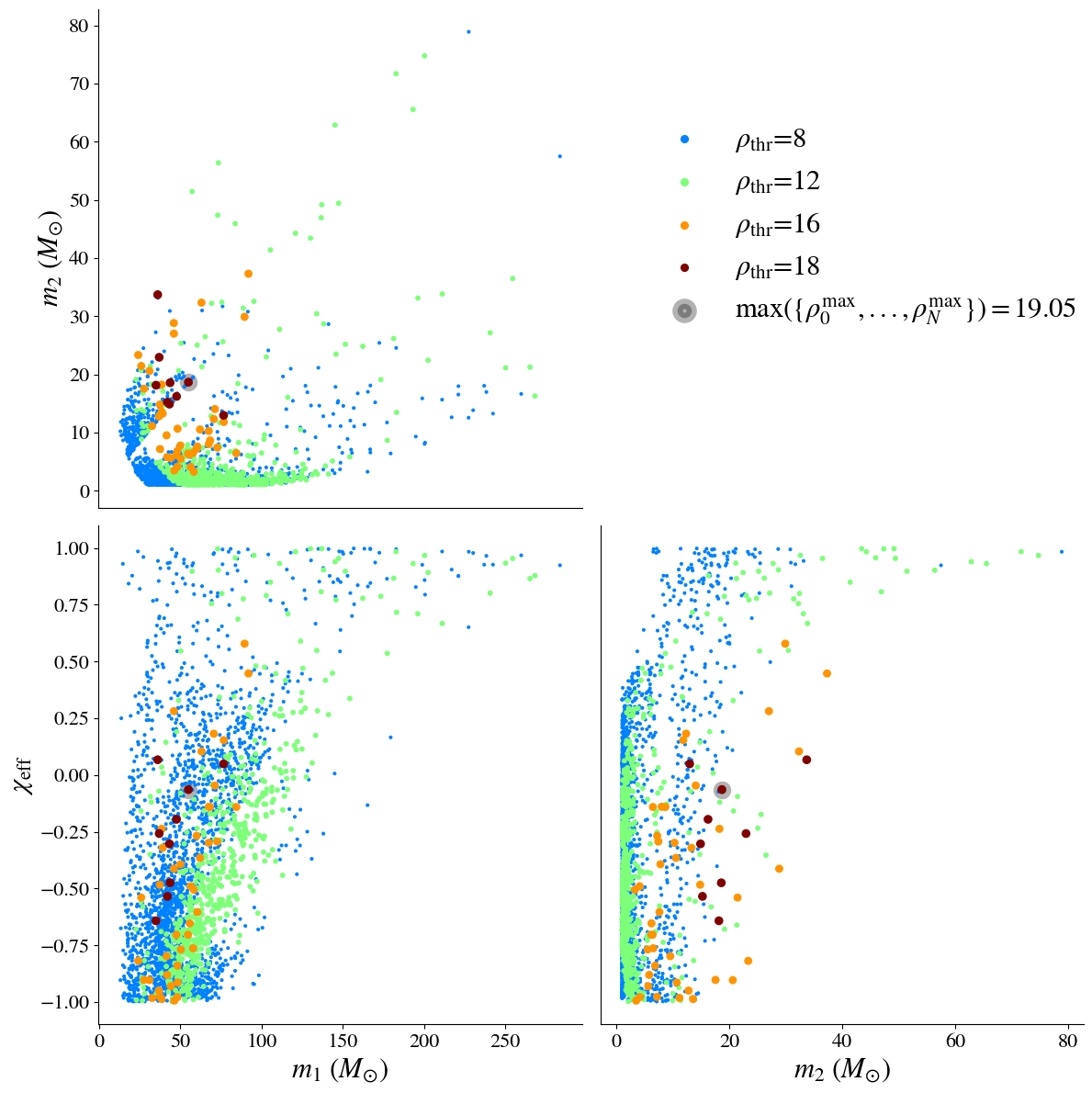}
	\caption{The positions of templates in the bank that have their corresponding states amplified after applying Grover's operator $k_{\ast}$ times to an initially equal superposition of template states for $\rho_{\text{thr}}$. Here $k_{\ast}$ is assumed to be the most probable $k_{\ast}$ from the outcome probabilities shown in Fig.~\ref{fig:psi2k}. The templates are scattered across the binary system's component masses $m_{1}$ and $m_{2}$ as well as the effective spin $\chi_{\textrm{eff}}$. The template marker size is proportional to the log probability of obtaining that template state from a measurement of the template register. With increasing $\rho_{\text{thr}}$ the matching templates cluster more tightly together and around the template found to have the maximum $\rho$ out of all the template (found from a classical search).}
	\label{fig:psiopt}
\end{figure*}

Fig.~\ref{fig:psiopt} shows the template states that are amplified from the Grover's operations in their corresponding positions in the parameter space for each of the different $\rho_{\text{thr}}$ cases from Sec.~\ref{sec:GWSD}. 
%For our example each measured $k_{\ast}$ is assumed to be the most probable value from the distributions shown in Fig.~\ref{fig:psi2k}.
The component masses $m_{1}$ and $m_{2}$ of each binary system are compared to the system's effective spin $\chi_{\textrm{eff}}=(s_{1}/m_{1}+s_{2}/m_{2})/(m_{1}+m_{2})$, a reparameterization of the component spins that adequately expresses their effect on the template waveforms in a single parameter. The colour of the template markers indicate the maximum $\rho_{\text{thr}}$ that correspond to them meeting the matching criteria. Note that all templates that correspond to a high $\rho_{\text{thr}}$ are a subset of lower $\rho_{\text{thr}}$ values, such that all templates plotted are matches for $\rho_{\text{thr}}=8$ but only those marked in red correspond to $\rho_{\text{thr}}=18$.
The size of the template labels is scaled to the log probability of obtaining the index of that template from the measurement (where each matching template is obtained with equal probability of $P(\text{Match})/r$) assuming the most probable $k_{\ast}$ Grover's operations from Fig~\ref{fig:psi2k} are applied. The classically calculated maximum $\rho$ across all the templates is found to be $19.05$ and is highlighted in the figure. 
This maximum $\rho$ template coincides with one of the templates that correspond to a match with $\rho_{\text{thr}}=18$.

\begin{figure}
	\centering
	\includegraphics[width=0.45\textwidth]{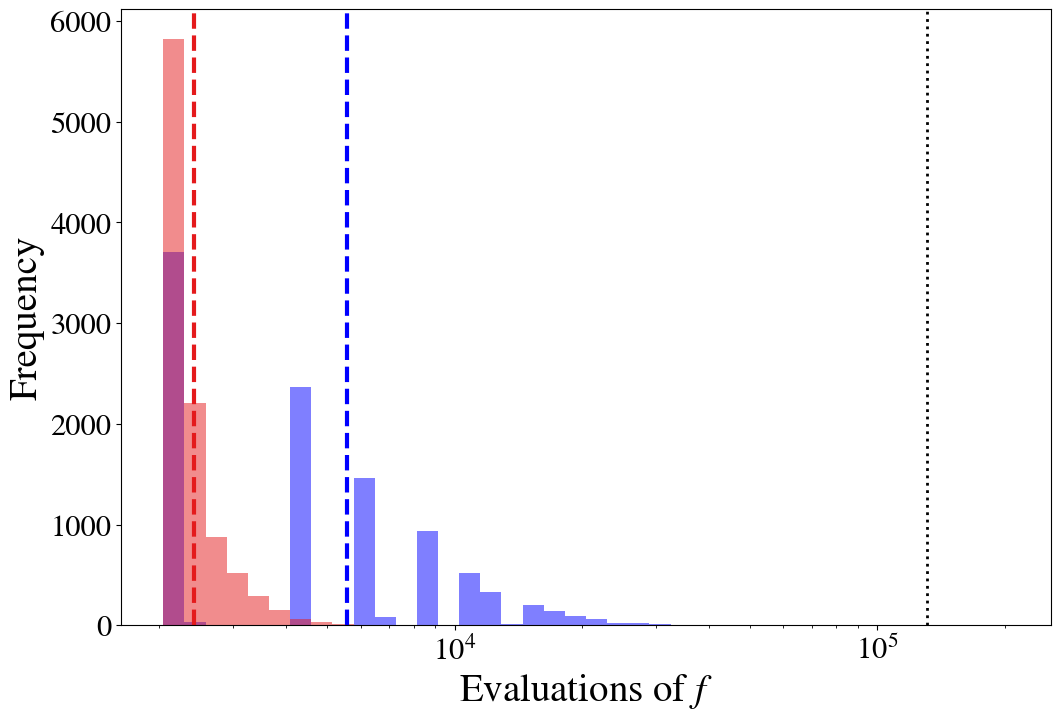}
	\caption{The number of evaluations of $f$ required to retrieve a matching template for 10,000 simulations given the GW150914 example with $\rho_{\text{thr}}=18$ and $p=11$. The red histogram corresponds to simulations where the value of $k_{\ast}$ from the \textsc{Signal Detection} procedure is assumed for \textsc{Template Retrieval}, which is repeated until a matching template is found. The blue histogram depicts simulations where the quantum counting algorithm is repeated to obtain a new $k_{\ast}$ for each application of the \textsc{Template Retrieval} algorithm. The mean for both extreme methods of $\sim2,418$ and $\sim5,575$ (red, blue dashed lines respectively) are compared to the classical case where all $2^{17}$ templates are evaluated (dotted line).}
	\label{fig:sims}
\end{figure}

It must be highlighted that non-optimal outcomes of measuring the counting register will often occur, and the corresponding $k_{\ast}$ used in the \textsc{Template Retrieval} procedure will not maximally amplify the matching template states and increase the probability of failing to retrieve a template as explored in Sec.~\ref{sec:RMTAlg}.
Even given $k_{\ast}=k$, there is a non-zero probability of failing to retrieve a template. Although repeating the algorithm if a match is not found does not add to the asymptotic complexity of the algorithm, which remains $O(\sqrt{N})$, we are also interested in the pre-factors, for a rigorous comparison between classical and quantum algorithms. In the remainder of this section we explore strategies to retrieve a template given a non-zero probability of failure, and benchmark these against the classical case.

If the \textsc{Template Retrieval} procedure fails to return a matching template then we can choose to repeat the algorithm given the same $k_{\ast}$ until a matching template is found. 
Given the $\rho_{\text{thr}}=18$ case with $p=11$, we carry out 10,000 simulations of measuring the counting register after the \textsc{Signal Detection} procedure to obtain $k_{\ast}$, before repeating \textsc{Template Retrieval} for each $k_{\ast}$ until a matched template is found.
%For each simulation the function $f$ is evaluated, defined as applying Eq.~\ref{equ:discretesnr} to the data via the \ac{FFT} and determining if the maximum $\rho$ in the resulting time-series is greater than $\rho_{\text{thr}}$. 
The number of times $f$ (from Eq.~\ref{equ:fsnr}) is evaluated for each simulation is tallied in the red histogram of Fig.~\ref{fig:sims} with a mean indicated by the red dashed line. 
This can be compared to the number of times $f$ is evaluated in the classical search case where the function is called upon for every template, indicated by the black dotted line.
An alternative approach after repeated failures to retrieve a matching template may be to assume the given $k_{\ast}$ is sub-optimal, and to reapply the \textsc{Signal Detection} procedure for another $k_{\ast}$ to use.
We caution that as the computational cost of the \textsc{Signal Detection} procedure is at least $\sim4$ times more costly than \textsc{Template Retrieval} the tolerance to the number of failed applications of \textsc{Template Retrieval} should be $\gg 1$.
To illustrate this, a further 10,000 simulations are made as before, but the \textsc{Signal Detection} procedure is repeated to give a new $k_{\ast}$ for each application of the \textsc{Template Retrieval} procedure, corresponding to a fail tolerance of 1.
The number of $f$ evaluations of these simulations using this extreme method is shown in the blue histogram of Fig.~\ref{fig:sims} and can be seen to have a much greater cost than the method without a fail tolerance.
The intervals between adjacent blue histogram bins correspond to the factors of $2^{p}-1$, the number of $f$ evaluations in applying \textsc{Signal Detection} to obtain the new $k_{\ast}$. 
Interestingly the mean number of $f$ evaluations for this extreme case is still significantly less than the classical case of calculating $f$ for all templates.
While some choice of failure tolerance may somewhat reduce the tail of the distribution above $\sim2(2^{p}-1)$, this corresponds to a fraction of $\sim0.01$ of the simulations when no failure tolerance is applied and is therefore insignificant for this case where $N/r=2^{17}/9$ and $p=11$.

For the case when all matching templates are desired, then the step described previously must be repeated as described in Step 6 of Sec.~\ref{sec:psuedocode}, which leads to matching templates being sampled with replacement.
This step would be costly for low $\rho_{\text{thr}}$ with a large proportion of matching templates, which may occur for a loud signal and a low $\rho_{\text{thr}}$ used for detection. 
However a procedure can be made using these algorithms as subroutines to obtain matches with a high $\rho$ while searching using a low $\rho_{\text{thr}}$; a low $\rho_{\text{thr}}$ can initially be assumed for the search specified in Sec.~\ref{sec:GWSD}, and given a measurement corresponding to $P(r_{\ast}>0)$, the value of $r_{\ast}$ obtained can be assessed.
If $r_{\ast}\gg1$, and the signal is presumed to be loud, then the steps in Sec.~\ref{sec:GWSD} can be repeated with larger $\rho_{\text{thr}}$. 
This can be repeated to optimize the choice of $\rho_{\text{thr}}$ until the desired number of templates is obtained.
The corresponding value of $k_{\ast}$ from this step can then be used to amplify the matching templates.
However, each step of this optimization approach requires applying the more computationally costly \textsc{Signal Detection} procedure and therefore should be made as to minimize the number of steps, which is a point for future work.

%%%%%%%%%%%%%%%%%%%%%%%%%%%%%%%%%%%%%%%%%%%%%%%%
%%%%%%%%%%%%%%%%%%%%%%%%%%%%%%%%%%%%%%%%%%%%%%%%
\section{Application: Continuous waves}\label{sec:cw_example}

The toy model example using Qiskit and the realistic practical example applied to the GW150914 data serve primarily as demonstrations of the method. The most impactful application of this algorithm for \ac{GW} data analysis is for problems where the optimal matched filtering approach is intractable via current classical computing. The continuous \ac{GW} case is such a problem due to the vast numbers of templates required to cover the search space for unknown continuous wave sources in order to perform a fully coherent search. A fully coherent search is one in which the match between template and data assumes phase coherence for the duration of the data span. Semi-coherent approaches use shorter data segments, requiring significantly less templates, and then incoherently combine results from each segment. This latter approach is computationally feasible but has reduced sensitivity.

If performing a fully coherent search for a continuous \ac{GW} signal the simplest model to assume for the time varying signal phase as defined at the \ac{SSB} can be further expressed as the Taylor expansion 
\begin{equation}
    \Phi(t_{\text{SSB}},\vec{\theta}) = \phi_{0} + 2\pi \sum_{k=1}\frac{f_{k}t_{\text{SSB}}^{k}}{k!}
\end{equation}
where $f_{k}$ being the $k$'th derivative of the phase with respect to the \ac{SSB} time. We further require the transformation between the times defined at the \ac{SSB} and the detector frame which we represent as  
\begin{equation}
t_{\text{SSB}} = t + \vec{r}(t)\cdot n(\alpha,\delta_{\text{d}}) + \delta t_{\text{parallax}} + \delta t_{\text{Shapiro}} + \delta t_{\text{Einstein}} + \delta t_{\text{binary}}. 
\end{equation}
The first term here (the Roemer delay) is the dominating contribution to the timing correction. This term is due to the varying position of the detector $\vec{r}(t)$ as the Earth spins and the orbits the Sun relative to the position of the source on the sky. We denoted the source position by the unit vector $n(\alpha,\delta_{\text{d}})$ dependent on the right ascension $\alpha$ and declination $\delta_{\text{d}}$. For observations of length $\sim 1$ year it is orbital motion in particular that then dictates the number and density of templates that are required on the sky parameters.

A rigorous calculation of the parameter-space metric governing the sky and the \ac{GW} frequency and its derivatives can be found in ~\cite{2007PhRvD..75b3004P} when applied to the so-called $\mathcal{F}$-statistic~\cite{1998PhRvD..58f3001J}. This statistic is the maximum likelihood ratio for a given template location analytically maximised over the 4 amplitude parameters (the received strain amplitude $h_{0}$, the initial reference phase $\phi_{0}$, the polarisation angle $\psi$, and the inclination angle $\iota$) governing a continuous signal. A useful approximation of the number of required templates can be obtained by considering the allowed variation in each of the search parameters that would lead to a 1 radian phase difference over the course of an observation. This is based on the fact that such a phase difference between signal and template would result in a tolerable level of \ac{SNR} loss for a coherent analysis. This order of magnitude calculation gives us
\begin{equation}
    N\sim 2\times 10^{28}\left(\frac{f}{1\text{kHz}}\right)^{2}\left(\frac{T}{1\text{year}}\right)^{3}\left(\frac{\Delta f}{1\text{Hz}}\right)\left(\frac{\Delta f_1}{10^{-9}\text{Hz}\,\text{s}^{-1}}\right)
\end{equation}
as the total number of templates to search the entire sky over a 1Hz frequency band at 1kHz. Typical searches are performed on small sub-bands analysed in parallel on $\sim 1000$ node computing clusters.

In a similar fashion to the technique used in the \ac{CBC} search to optimise the search over time of arrival, the \ac{FFT} can be used to optimise the search over the intrinsic frequency $f_{0}$. Hence the template bank can be divided into the Cartesian product between frequency templates and the remainder, where the overall classical search cost is linear in the number of templates over the sky and frequency derivative:
\begin{equation}
    N_{\text{sky},f_{1}} \sim 10^{20}\left(\frac{f}{1\text{kHz}}\right)^{2}\left(\frac{T}{1\text{year}}\right)^{2}\left(\frac{\Delta f_1}{10^{-9}\text{Hz}\,\text{s}^{-1}}\right),
\end{equation}
but the joint cost of calculating the detection statistic for a single sky and frequency derivative value, over all possible intrinsic frequencies scales as $N_{f_{0}}\log{N_{f_0}}$ where
\begin{equation}
    N_{f_{0}} \sim 2\times 10^{8}\left(\frac{T}{1\text{year}}\right).
\end{equation}
The total number of templates in this simple scenario, even when considering a narrow band 1Hz search is many orders of magnitude greater than the number searched in previous analyses (in~\cite{2019ApJ...875..122A} the total number of templates searched was $\sim 10^{14}$ which also included templates over the 2nd frequency derivative $f_{2}$). Hence, the fully coherent all-sky search over frequency and frequency derivative for 1 year of data is currently completely infeasible using classical computing. 

We have shown that the quantum approach offers a speed-up of $O(\sqrt{N_{\text{sky},f_{1}}})$ in the number of calculations required. However, the big O notation refers to asymptotic scaling, and tells us nothing about the pre-factors, which could be different in the classical and quantum cases. To claim an expected improvement for a particular case we need to be a bit more precise. To be specific, for the calculation of the detection statistic, the quantum algorithm requires precisely the same steps as the classical algorithm, but requires these to be done in a reversible way, and in addition requires the reversal of the calculation to be performed each time, in order to disentangle these registers from the index register. Standard techniques may be used to construct reversible versions of classical Boolean circuits, which may be implemented directly as quantum circuits. Any classical circuit with $T$ gates and $S$ bits may be converted to a reversible circuit with $O(T^{1+\Delta})$ gates and $O(S \log T)$ bits. Specifically, for any $\Delta >0$ it is possible to construct a reversible circuit in which the number of gates required is upper bounded by $3T^{1+\Delta}$~\cite{rieffel2011}. We thus neglect the factor $T^\Delta$, which may be made arbitrarily small, leading to a factor of $3$ in the number of gates required. The requirement to erase the intermediate calculations adds a further factor of two, thus there is a factor of 6 in the number of gates required for the detection statistic calculation in the quantum algorithm compared to the classical algorithm.

Classically, to be certain there is no signal, we need to check against all templates, so we require $N_{\text{sky},f_{1}}$ such calculations. In the quantum algorithm to determine whether there is a match or not, we choose $p$ to be the smallest integer larger than $p = \log (\pi \sqrt{N})$, requiring around $\pi \times 10^{10}$ iterations. This gives a false negative with probability at most $1/\pi^2$. As discussed, $\ell$ repetitions of the whole procedure reduce the probability of a false negative to $\pi^{-2\ell}$. Thus, e.g. a false negative probability of order $10^{-6}$ requires $6$ repetitions. Finally, the inverse Fourier transform and measurement steps result in an addition of a logarithmic number of gates, and may be neglected. Overall, for a false negative probability of $10^{-6}$ we therefore require around $2 \times 10^{11}$ iterations, each of which requires a factor of $6$ more gates than the classical calculation. The overall number of gates needed is of order $10^{12} T$, compared to $10^{20} T$ classically, representing a reduction by a factor of $10^{8}$ in the number of operations required.

%Discuss things about practicalities for the future. How many qubits are needed
%etc.. 

%\begin{itemize}
%\item space requirements on quantum computer
%\item describe the current state of the art in quantum processors
%\item run on the actual quantum processor and compare the results, efficiency and gate costs
%\end{itemize}

%\cm{I agree with all of this but there needs to be text added here.}

%%%%%%%%%%%%%%%%%%%%%%%%%%%%%%%%%%%%%%%%%%%%%%%%
%%%%%%%%%%%%%%%%%%%%%%%%%%%%%%%%%%%%%%%%%%%%%%%%
\section{Discussion}\label{sec:discussion}
We have presented a quantum algorithm for matched filtering for gravitational wave data analysis. Our algorithm, based on Grover's search algorithm, offers a square-root speed-up in the computational cost of searching through a large template bank. As the number of templates is the limiting factor regarding computational feasibility in gravitational wave analysis for certain astrophysical signals, this is a natural application of Grover's algorithm. The key theoretical insight that we have used is that for problems of astrophysical interest, the templates are readily computable from theoretical models, and need not be pre-stored in a database, thus eliminating the need for qRAM. This allows us to construct an oracle, which is readily used in Grover's algorithm, and its extension in quantum counting, to determine whether there are templates that produce an \ac{SNR} above a given threshold, and to find matching templates.

We have presented proof-of-principle demonstrations of template matching on IBM Qiskit, and through a python simulation applied to actual gravitational wave data. We have also discussed the application to continuous wave searches, currently infeasible with classical techniques. We have focussed on applications to gravitational wave data analysis, but the algorithm presented here could of course be readily applied to any template matching problem in which the number of templates is much bigger than the size of any one template, and in which the templates can be calculated efficiently. %We note that many existing applications of quantum algorithms in physics are quantum simulations, taking advantage of better simulation of physical systems by quantum processors \cite{georgescu2014quantum}. This algorithm is of note, perhaps, as its primary use will be in studying physical systems, but it is not at heart a quantum simulation. \smc{I don't think I can stand by the last couple of sentences any more: the recent particle physics papers that we quote earlier are also of this flavour, and it occurs to me that variational quantum optimisers, of wide use in quantum chemistry, are not really quantum simulations either. I would suggest to just get rid of the last two sentences, unless there are any objections (we discussed this recently and it was decided to keep it but rephrase, but this was for reasons of clarity, not accuracy).}

As we are still some way from scalable, error-corrected quantum processors, it is worth outlining the space requirements of our algorithm, as well as the gate complexity. With $N$ templates and signal data consisting of $M$ time steps, we require a counting register of size $\lceil \log_2 \pi + \frac{1}{2} \log_2 N \rceil$ qubits, an index register of $\log_2 N$ qubits, and two registers of $64 M$ qubits (if each time sample is stored as a floating point number, using 8 bytes, or 64 classical bits): one to store the data, and one to store a template corresponding to each index. Recall that these are stored in superposition, so only one template register is needed. In addition, to produce the templates and perform the matched filtering calculation reversibly, we introduce a modest space overhead logarithmic in $M$. The dominant contribution to the overall space needed is therefore the size of the data. For the example given in section \ref{sec:cbcexample}, this is 28 seconds of data at 4096Hz, giving $M = 28 \times 4096$. With 8 bytes for each data point, our algorithm becomes feasible with an error-corrected device with a few Megabytes of memory. For fully coherent searches over longer datasets, this increases linearly, and the continuous wave application discussed in section \ref{sec:cw_example} requires around $3{\rm Gb}$ of memory. The current state-of-the-art is around 50-100 physical qubits \cite{arute2019quantum,IBMRoadmap}. Nonetheless, IBM's ambitious quantum hardware roadmap aims for over 1000 qubits by 2023, in their proposed Condor processor, a device that they view as ``a milestone that marks our ability to implement error correction and scale up our devices" \cite{IBMRoadmap}. 

We note also that we have discussed so far only the gate complexity. In the first error-corrected devices, quantum gate operations will be much slower than their classical counterparts, due to both intrinsic gate operation times and the overhead introduced by quantum error-correction. Quadratic speed-ups, such as that discussed here, do not seem to be promising for runtime advantages for modest fault-tolerant devices \cite{PRXQuantum.2.010103}. Taking this into account, combined with the quite demanding space requirements outlined above, we do not claim this as a near term application. However, in the medium to long term with improvements in quantum hardware and in error correction, quantum algorithms have the potential to offer significantly improved sensitivity in gravitational wave searches.

This represents just the first step in constructing possible applications of quantum computation to gravitational wave data analysis. Employing Grover's algorithm to speed-up the search for a match within a large template bank is the first natural step in exploring connections between the two fields. Possibilities for improvement could be to incorporate prior knowledge into the initial state prepared, giving higher weighting to templates considered a priori more likely. This has already been explored classically \cite{2010JPhCS.228a2008R,Dent:2013cva}, and as long as the resulting superposition may be prepared efficiently, such approaches remain amenable to amplitude amplification \cite{brassard2002quantum}. The speed-up relative to the classical case would remain quadratic, but the overall efficiency of both algorithms can be improved. 

Another possibility is employing amplitude encoding to store the data and templates. In amplitude encoding the amplitude of the data at a given time point is stored as the amplitude of a quantum state. This would significantly reduce the space requirements, from an $O(M)$ qubit processor to $O(\log M)$. The advantage of the digital encoding we have used here is that arithmetic operations, needed to produce the templates and compute the SNR to check for a match above threshold are readily translated from classical circuits. The required matching is more challenging using amplitude encoding, and would likely add to the complexity of this step.
%There is already a very broad literature in quantum machine learning and related areas with applications to data processing, and it is worth discussing why we suggest that Grover's algorithm, with its modest square-root speed-up is the natural approach for matched filtering. Algorithms for quantum template matching were first proposed almost 20 years ago~\cite{sasaki2001,sasaki2002}. These provide a fully quantum analogue of learning tasks: both the data and templates are provided as quantum states, and optimal strategies for determining the closest matching template are given. A related task in the literature is estimating the overlap between quantum states, provided a number of copies of each~\cite{buhrman2001,fanizza2020}. Such techniques are useful when the data and templates are provided as quantum states, and further, when the dimension spanned by the states \sjg{what does the `states' refer to here? the size of the data or template or the number of templates?} is the limiting parameter. In the case of interest here the data is classical, and more importantly the number of templates is by far the largest parameter, and quantum template matching algorithms do not obviously offer an advantage.
A final possibility is to apply machine learning techniques, either in digital or amplitude encoding, to analyse gravitational wave data. This seems promising as machine learning is considered a promising area of study for applications for NISQ devices~\cite{Preskill2018}. Classical machine learning techniques are beginning to be employed in gravitational wave detection~\cite{gabbard2018matching,2019arXiv190906296G} as well as other gravitational wave areas~\cite{cuoco2020enhancing}, and we expect that more sophisticated quantum machine learning techniques may yield further quantum advantages. Exploring the possibility of amplitude encoding, and of quantum machine learning are however left for future work. As we fully enter the era of gravitational wave astronomy, better performing and more efficient data processing techniques will be needed to fully exploit this new window on the Universe. In parallel, as we embark on an era of quantum computational advantage, we anticipate a fruitful interplay between the two fields in harnessing the new computational capabilities offered by this emerging technology.

%%%%%%%%%%%%%%%%%%%%%%%%%%%%%%%%%%%%%%%%%%%%%%%%
\begin{acknowledgments}
The authors also gratefully acknowledge the Science and Technology Facilities Council of the United Kingdom. CM and JV are supported by the Science and Technology Facilities Council (grant No. ST/ L000946/1). FH is supported by STFC grant number ST/N504075/1.
\end{acknowledgments}

%%%%%%%%%%%%%%%%%%%%%%%%%%%%%%%%%%%%%%%%%%%%%%%%
\appendix

%%%%%%%%%%%%%%%%%%%%%%%%%%%%%%%%%%%%%%%%%%%%%%%%
%%%%%%%%%%%%%%%%%%%%%%%%%%%%%%%%%%%%%%%%%%%%%%%%
\section{Variable Summary}\label{sec:variableSum}
Table~\ref{tab:nomen} is a summary of the various variables used in the paper.

\begin{table}[h!]	
\resizebox{\columnwidth}{!}{%
\begin{tabular}{c|c}
\hline
\hline
\hspace{0.1cm}Variable\hspace{0.1cm} & Description \\
\hline
$r$ & True number of matching templates\\
$\tilde{b}$ & the non-integer counting register value corresponding to $r$\\
$k$ & Number of Grover operations corresponding to $r$\\
$2\theta$ & Rotation in state space corresponding to $r$\\
\hline
$b$ & Observed counting register outcome\\
$r_{\ast}$ & Number of matching templates corresponding to $b$\\
$k_{\ast}$ & Number of Grover operations corresponding to $r_{\ast}$\\
$2\theta_{\ast}$ & Rotation in state space corresponding to $r_{\ast}$\\
\hline
\end{tabular}}
\caption{Nomenclature used throughout the text where inferred variables are denoted by the `$\ast$' subscript. \label{tab:nomen}}
\end{table}

\section{Quantum Gates}\label{sec:quantumgates}
A quantum computer is roughly composed of three parts: 1) Quantum registers to store qubits; 2) A series of quantum gates to perform unitary transformations on the input states; and 3) the measurement procedure to readout the final result.

The qubits have only two orthogonal states, similar to classical computation. The computational basis states are labeled by the associated binary string. They are often represented by column vectors as: 
\begin{equation}
    \label{equ:quantumstate}
  %\begin{align}
    \ket{0} = \begin{bmatrix}
           1 \\
           0 
         \end{bmatrix},
         \quad
         \ket{1} = \begin{bmatrix}
           0 \\
           1 
         \end{bmatrix}.
  %\end{align}
\end{equation}

The other pair of orthogonal states frequently used are $\ket{+}$ and $\ket{-}$, defined as:
\begin{equation}
    \label{equ:quantumstate+}
  %\begin{align}
    \ket{+} = \frac{1}{\sqrt{2}}(\ket{0}+\ket{1})=\frac{1}{\sqrt{2}} \begin{bmatrix}
           1 \\
           1 
         \end{bmatrix},\\
         \ket{-} = \frac{1}{\sqrt{2}}(\ket{0}-\ket{1})=\frac{1}{\sqrt{2}}\begin{bmatrix}
           1 \\
           -1 
         \end{bmatrix}.
  %\end{align}
\end{equation}

Quantum gates are normally represented by unitary matrices. The quantum gates only applied to one qubit are called single-qubit gates and the ones involve multiple qubits are called multiple-qubit gates. 

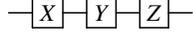
\begin{figure}[h]
\[\Qcircuit @C=1em @R=.7em {
   & \gate{X} & \gate{Y} & \gate{Z} & \qw 
}\]
  \caption{\label{fig:pauli}
The Pauli gates expressed in a quantum circuit.}
\end{figure}

One set of the most frequently used single-qubit gates are the Pauli gates, whose matrix forms are the associated Pauli matrices. They rotate the qubit by $\pi$ radiance around the corresponding axis on the Bloch sphere. The Pauli-X operator is particular of interest, because it functions as the classical NOT gate. They are represented in  a quantum circuit diagram shown in Fig.~\ref{fig:pauli}.

\begin{figure}[h]
\[\Qcircuit @C=1em @R=.7em {
    & \gate{H} & \qw 
}\]
  \caption{\label{fig:hadamard}
The Hadamard gate expressed in a quantum circuit.}
\end{figure}
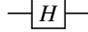

Another important single qubit gate is the Hadamard gate, which interchange the states between the computational basis and the $\ket{+}$ and $\ket{-}$ basis:
\begin{equation}
    \label{equ:hadamard}
  %\begin{align}
    \hat{H} = \frac{1}{\sqrt{2}}\begin{bmatrix}
           1& 1\\
           1 &-1
         \end{bmatrix},
 % \end{align}
\end{equation}
and represented in a quantum circuit as shown in Fig.~\ref{fig:hadamard}

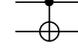
\begin{figure}[h]
\[\Qcircuit @C=1em @R=.7em {
    &\ctrl{1}&\qw\\
    & \targ & \qw 
}\]
\caption{\label{fig:cnot}The CNOT gate expressed in a quantum circuit.}
\end{figure}

The multiple-qubit gates used in this paper are controlled gates which is often written as $C^n$-$U$. A controlled gate act on the state of two types of qubits: the control qubits and the target qubits. The operation will be applied to the target qubit if and only if all the $n$ control qubits are in state $\ket{1}$. One example would be the CNOT gate:
\begin{equation}
    \label{equ:cnot}
  %\begin{align}
    \hat{U}_{\text{CNOT}} = \begin{bmatrix}
           1& 0&0&0\\
           0 &1&0&0\\
           0&0&0&1\\
           0&0&1&0
         \end{bmatrix},
 % \end{align}
\end{equation}
and its corresponding quantum circuit expression is shown in Fig.~\ref{fig:cnot}.

\section{Probability of false negative}\label{sec:Pt0}
Recall the state of the register after inverse Fourier transform $\ket{\psi_6}$ in Eq.~(\ref{equ:InverQFT}). Without losing generality, only one eigenstate is considered for the analysis. According to Eq.~(\ref{equ:InverQFT}), the probability of a certain $\ket{b}$ is measured in the whole state would simply be twice of the probability of that in one eigenstate. To consider the amplitude for the measured state $\ket{b}$ for eigenstate $\ket{s_+}$, we can sum up all its amplitude across $a$:
%\begin{equation}
  \begin{align}
    \label{equ: amplitude}
  &\frac{1}{2^{p}}\sum_{a=0}^{2^p-1}e^{i2\pi a(\frac{\theta}{\pi}-\frac{b}{2^p})}\ket{b}= \frac{1}{2^{p}}\frac{e^{i2\pi 2^p(\frac{\theta}{\pi}-\frac{b}{2^p})}-1}{e^{i2\pi (\frac{\theta}{\pi}-\frac{b}{2^p})}-1}\ket{b}\nonumber\\
  &=\frac{1}{2^{p}}\frac{\sin\left(\pi 2^p(\frac{\theta}{\pi}-\frac{b}{2^p})\right)}{\sin\left(\pi (\frac{\theta}{\pi}-\frac{b}{2^p})\right)}e^{i\pi (2^{p}-1)(\frac{\theta}{\pi}-\frac{b}{2^p})}\ket{b}.
\end{align}  
%\end{equation}

The probability of state $\ket{b}$ would be:
\begin{equation}
    \label{equ:probabilityOG}
P(b)= \frac{1}{2^{2p}}\bigg(\frac{\sin\big(2^p \theta\big)}{\sin (\theta-\frac{b\pi}{2^p})}\bigg)^2.
\end{equation}
From the discussion in Sec.~\ref{sec:qmf}, the only state situation will trigger a no signal result is when $\ket{b}=0$. According to Eq.~\ref{equ:probabilityOG}, the probability of false negative is:
\begin{equation}
  \begin{aligned}
    \label{equ:falneg1}
    P(r_{\ast}=0|r>0)&=P(b=0)\\
    &=\frac{1}{2^{2p}}\Big(\frac{\sin\big(2^p\theta\big)}{\sin (\theta)}\Big)^2.
    \end{aligned}
\end{equation}

\bibliography{main}

\end{document}